\documentclass{aa}   

\usepackage{graphicx}
\usepackage{txfonts}
\usepackage{natbib}  
\usepackage[figuresright]{rotating}
\usepackage{color}

\begin{document}

\title{CSI2264: Simultaneous optical and X-ray variability in the pre-main sequence stars of NGC~2264}
\subtitle{II: Photometric variability, magnetic activity, and rotation in class~III objects and stars with transition disks.}

\author{M. G. Guarcello\inst{1} \and E. Flaccomio\inst{1} \and G. Micela\inst{1} \and C. Argiroffi\inst{1,2} \and S. Sciortino\inst{1} \and L. Venuti\inst{3} \and J. Stauffer\inst{4} \and L. Rebull\inst{4} \and A. M. Cody\inst{3,5}
          }

\institute{INAF - Osservatorio Astronomico di Palermo, Piazza del Parlamento 1, I-90134, Palermo, Italy\\
              \email{mario.guarcello@inaf.it}
          \and
              Dip. di Fisica e Chimica, Universit\`{a} di Palermo, Piazza del Parlamento 1, 90134 Palermo, Italy 
          \and
              NASA Ames Research Center, Moffett Field, CA 94035, USA
          \and
              Spitzer Science Center, California Institute of Technology, Pasadena, CA 91125, USA 
          \and
              Bay Area Environmental Research Institute, 625, 2$^{nd}$ St Ste. 209, Petaluma, CA 94952
              }

\date{}

\abstract
{Pre-main sequence stars are variable sources. In diskless stars this variability is mainly due to the rotational modulation of dark photospheric spots and active regions, as in main sequence stars even if associated with a stronger magnetic activity.}
{We aim at analyzing the simultaneous optical and X-ray variability in these stars to unveil how the activity in the photosphere is connected with that in the corona, to identify the dominant surface magnetic activity, and to correlate our results with stellar properties, such as rotation and mass.}
{We analyzed the simultaneous optical and X-ray variability in stars without inner disks (e.g., class III objects and stars with transition disks) in NGC~2264 from observations obtained with $Chandra$/ACIS-I and CoRoT as part of the Coordinated Synoptic Investigation of NGC~2264. We searched for those stars whose optical and X-ray variability is correlated, anti-correlated, or not correlated by sampling their optical and X-ray light curves in suitable time intervals and studying the correlation between the flux observed in optical and in X-rays. We then studied how this classification is related with stellar properties.}
{Starting from a sample of 74 class III/transition disk (TD) stars observed with CoRoT and detected with Chandra with more than 60 counts, we selected 16 stars whose optical and X-ray variability is anti-correlated, 11 correlated, and 17 where there is no correlation. The remaining stars did not fall in any of these groups. We interpreted the anti-correlated optical and X-ray variability as typical of spot-dominated sources, due to the rotational modulation of photospheric spots spatially coincident to coronal active regions, and correlated variability typical of faculae-dominated sources, where the brightening due to faculae is dominant over the darkening due to spots.} 
{Stars with ``anti-correlated'' variability rotate slower and are less massive than those with ``correlated'' variability. Furthermore, cool stars in our sample have larger $u-r$ variability than hot stars. This suggests that there is a connection between stellar rotation, mass, and the dominant surface magnetic activity, which may be related with the topology of the large-scale magnetic field. We thus discuss this scenario in the framework of the complex magnetic properties of weak-line T Tauri stars observed as part of recent projects.}
 

\maketitle

\section{Introduction}
\label{intro}

The variability of pre-main sequence (PMS) stars has been investigated as part of several projects. One such project is the Young Stellar Object VARiability \citep[YSOVAR;][]{Rebull2014AJ.148.92R}, which explored the variability of young stellar objects in the mid-infrared using Spitzer Space Telescope \citep{Werner2004ApJS.154.1W} Infrared Array Camera \citep[IRAC;][]{Fazio2004ApJS.154.10F} observations of several clusters, such as the Orion Nebula Cluster \citep{Morales-CalderonSHG2011}, NGC 1333 \citep{Rebull2015AJ.150.175R}, and Serpens \citep{Wolk2018AJ.155.99W}. Another, related program is the Coordinated Synoptic Investigation of NGC~2264  \citep[CSI2264,][]{CodySBM2014AJ}, which monitored the young cluster NGC~2264 with 15 ground- and space-based telescopes simultaneously with the observations of the Convection, Rotation, and Planetary Transits  satellite \citep[\emph{CoRoT},][]{BaglinABD2006ESASP}. There have also been several monitoring projects using K2 \citep{Howell2014PASP.126.398H}: Upper Sco and $\rho$ Oph were monitored \citep{CodyHillenbrand2018AJ.156.71C,Rebull2018AJ.155.196R}, as were other young clusters. These projects have revealed the efficiency of multiwavelength time-domain astronomy as a powerful tool to study young stars, and have shown it to be capable of probing their stellar activity, accretion, and inner (a few 0.1$\,$AU) circumstellar environment in ways that greatly improve upon what can be done from the ground. \par

   The origin of variability in stars without inner dust disks (class III objects and stars with transition disks) is different from that in those with inner disks.  In class I/II sources, variability is dominated by disk-related phenomena, such as steady \citep[e.g.,][]{CemeljicSC2013} and unsteady \citep[e.g.,][]{RomanovaUKL2012,StaufferCBA2014} accretion, and variable extinction \citep{BouvierCAC1999,StaufferCMR2015AJ}. In disk-free stars, variability instead probes stellar rotation and magnetic activity occurring in the photosphere, chromosphere, and corona. Flares in these young stars are among the most powerful magnetic phenomena occurring in stars \citep[e.g.,][]{FletcherDHJK2011SSRv}. Photospheric spots, faculae, prominences, and coronal active regions contribute to stellar variability both because they are intrinsically variable phenomena, and because, as they rotate into and out of view, rotational modulation can be seen in the light curve. \citep{Grankin2008AA.479.827G}. \par
    
Photospheric spots and faculae are observed in great detail in the Sun. However, the magnetic activity in PMS stars is orders of magnitude more intense than in main sequence stars \citep[e.g.,][]{FeigelsonDecampli1981,Montmerle1996} and thus PMS stars are expected to be on average richer in photospheric and coronal active regions than the Sun. In some particularly active PMS stars, the presence of stellar spots distributed across most of the photosphere was deduced by Doppler images \citep[e.g.,][]{StrassmeierRice2006}, photometric variability \citep{Strassmeier2009AARv.17.251S}, and spectral fits to high-resolution NIR spectra adopting two thermal components accounting for quiet photosphere and cold spots \citep{Gully-Santiago2017ApJ.836.200G}. Similarly, the presence of coronal active regions is confirmed by several observations and simulations of the structure of the stellar coronae and their X-ray emission \citep[e.g.,][]{CohenDKH2010}. In addition, it has been suggested \citep[e.g.,][]{JardineCDG2006,Hill2019MNRAS.484.5810H} that PMS stars may have magnetic fields with complex topology, which results in highly structured stellar coronae and a rich population of photospheric and coronal magnetically active regions. Even at the age of the Pleiades, recent studies suggest spot coverage larger than 50\% in several late-type dwarf stars \citep{Fang2016MNRAS.463.2494F,JacksonJeffries2018MNRAS.476.3245J}. \par

    While rotational modulation in optical bands is widely used to study stellar rotation, X-ray rotational modulation due to coronal active regions in PMS stars has been observed in only a few cases, such as by \citet[][]{FlaccomioMSF2005ApJ} as part of the {\it Chandra Orion Ultradeep Project} \citep[COUP,][]{GetmanFBG2005}, a $\sim$13-day continuous $Chandra$/ACIS-I observation of the Orion Nebula Cluster ($\sim$1$\,$Myr). The COUP data, together with simultaneous optical observations in BVRI bands taken with the WIYN 0.9$\,$m telescope at the Kitt Peak National Observatory (KPNO) in Arizona, USA, and the 1.5$\,$m Cassini telescope in Loiano, Italy, were explored by \citet{StassunBFF2006} to search for correlated variability in optical and X-rays. They found correlated and anti-correlated optical and X-ray variability only in 5\% of the observed stars, concluding that in most PMS stars variability is not dominated by the simultaneous emergence of coronal and photospheric active regions or accretion hot spots during stellar rotation. However, their data were sparse and discontinuous (on average 8 data points per night). \par
     
        \begin{figure*}[]
        \centering      
        \includegraphics[width=13cm]{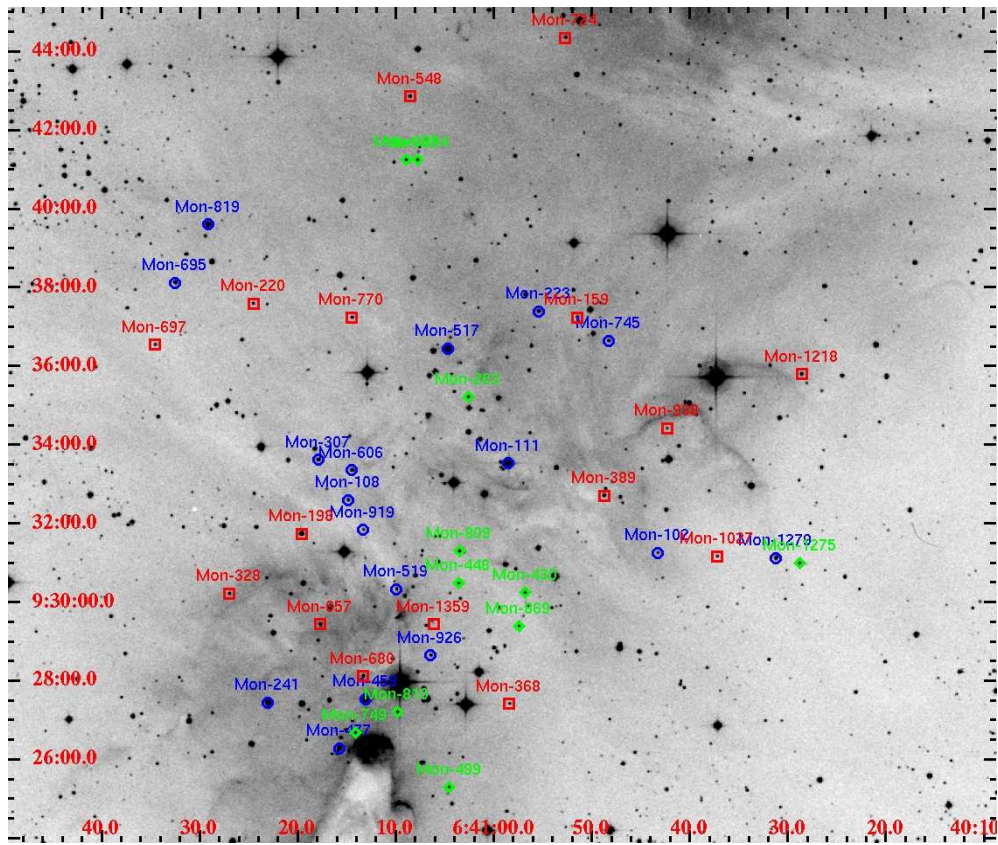}
        \caption{DSS-2 image of the central region of NGC~2264, with the sources analyzed in this paper marked with different colors and symbols corresponding to their final classification (red squares: anti-correlated sources; green diamonds: ``correlated'' sources; blue circles not correlated sources; see Sect. \ref{class_sec}). The labels indicate the Mon- name of each star. Right ascension in the x-axis, declination in the y-axis.}
        \label{field_img}
        \end{figure*}
    
    In this paper we study the simultaneous optical and X-ray variability of stars in NGC~2264. Figure \ref{field_img} shows a DSS-2 image of the central region of NGC~2264, with colored symbols indicating the position of the sources discussed in the present paper. The labels indicate the individual CSI MON- names (hereafter Mon-), which are the stellar identifiers adopted by the CSI2264 project. NGC~2264 is a young cluster \citep[1-5 Myrs,][]{RebullMSH2002,Dahm2008} that is relatively near to our Sun \citep[$760\,$pc,][]{ParkSBK2000}, being part of the local spiral arm. The cluster is characterized by low extinction, with a median value of $A_V=0.45^m$ \citep{RebullMSH2002}. Thanks to its proximity and small average extinction, its low-mass PMS stellar population is relatively easy to observe and analyze. The stellar population of this cluster is therefore well characterized, with a few early-type members identified, such as the O7V star S~Monocerotis \citep{SchwartzTOG1985}, about a dozen B type stars, and a rich low-mass population. NGC~2264 is the only cluster, together with the Orion Nebula Cluster, with such a large mass spectrum within 1 kpc of the Sun. In \citet[][]{Guarcello2017AA.602A.10G}, we analyzed the simultaneous variability in the optical and X-rays in stars with disks in the young open cluster NGC~2264 as part of the CSI~2264 project. As a follow-up to \citet{Guarcello2017AA.602A.10G}, which was focused on class~II objects, in this paper we take full advantage of the CSI~2264 long and uninterrupted monitoring observations to extend the analysis to stars with no inner circumstellar material, that is, class~III objects and stars with transitional disks that do not show disk-related variability. In particular, we study the correlation between simultaneous optical and X-ray variability. \par 
    
The paper is organized as follows: in Sect. \ref{data_sec} we describe the data analyzed in this paper and the sample of sources; in Sect. \ref{III_var} we identify the stars with correlated, uncorrelated, and anti-correlated X-ray and optical variability; in Sect. \ref{propert_sect} we study the properties of the stars with different correlations; in Sect. \ref{discuss_sect} we interpret our results and derive a possible scenario to explain the observed behavior. \par

\section{Targets and data}
\label{data_sec}

    In this paper, we analyze the broad optical band (``white'') CoRoT light curves, taken with a cadence of 32$\,$s or 512$\,$s. These light curves have been reduced using the standard CoRoT reduction pipeline \citep{SamadiFCD2006} which corrects from gain and zero offset, jitter, and electromagnetic interference, and also performs the background subtraction, and removes data resulting from hot pixels. We did not correct for those systematic errors that are not corrected by the standard pipeline, such as rapid flux jumps in the light curves due to rapid changes of the detector temperature. We simply ascertained that our results were not affected by the presence of such jumps. We used CoRoT data which are not flagged as suspicious data points.  \par

The X-ray data analyzed in this work were obtained from six $Chandra$/ACIS-I observations (see Table \ref{obsid_table}): Observations 9768 and 9769 were taken with a nominal exposure of 30$\,$ksec and co-pointed at $\alpha$=06:41:12, $\delta$=+09:30:00, while observations 13610, 13611, 14368, and 14369 were taken with a total nominal exposure of $300\,$ksec, and pointed  at $\alpha$=06:40:58.70, $\delta$=+09:34:14 (P.I. G. Micela). These Chandra/ACIS-I pointings were taken during the 2008 and 2011 CoRoT observations of NGC~2264. A detailed analysis of these X-ray data, including the source-detection process, photon extraction, and spectral fitting procedure, will be provided by Flaccomio et al. (in preparation). Briefly, all events were fully reprocessed using the CIAO task \emph{chandra-repro}. Sources were detected using the wavelet-based algorithm PWDetect \citep{Damiani1997ApJ.483.350D}, adopting a significance threshold of 4.4$\sigma$, roughly resulting in ten expected background fluctuations selected as sources. Event extraction, source repositioning, and source validation were performed with the IDL software $ACIS$ $Extract$ \citep[AE, ][]{BroosTFG2010}. After excluding candidate spurious detections, a total of 744 X-ray sources were validated. \par

    \begin{table}
    \caption{$Chandra$/ACIS-I observations log.}
    \label{obsid_table}
    \centering                       
    \begin{tabular}{ccc}        
    \hline\hline                
    Obs.ID & Exposure (ksec) & Date \\
    \hline                        
    9768  & 27.79 & 2008-03-12 \\
    9769  & 29.76 & 2008-03-28 \\
    14368 & 74.44 & 2011-12-3\\ 
    13610 & 92.54 & 2011-12-5\\
    13611 & 60.23 & 2011-12-7\\
    14369 & 66.16 & 2011-12-11\\
    \hline                                  
    \end{tabular}
    \end{table}

    The list of the CoRoT targets in the field of NGC~2264 includes 1617 known candidate members which were selected, as described by \citet{CodySBM2014AJ}, combining membership criteria based on: 1) the compatibility of stellar optical photometry with the cluster loci in color-magnitude diagrams \citep{FlaccomioMS2006}; 2) strong H$\alpha$ emission \citep{RebullMSH2002,LammBMH2004,SungBCK2008}; 3) X-ray detection \citep{RamirezRSH2004,FlaccomioMS2006}; 4) radial velocity \citep{FureszHSR2006}; and 5) the presence of a circumstellar disk \citep{SungSB2009}. From this list, we extracted the good candidate class~III objects as those stars not showing any evidence of the presence of a dust disk (e.g., no IR excesses). We also selected candidate nonaccreting stars with TDs as those with excesses only at 8.0$\,\mu$m and/or 24$\,\mu$m and no signatures of accretion (the latter requirement is imposed in order to avoid stars whose variability is affected by accretion\footnote{Accreting stars are selected from: the IPHAS $r-i$ vs. $r-H\alpha$ diagram \citep{LammBMH2004} and H$\alpha$ equivalent width larger than 10$\AA$ \citep[e.g.,][]{Venuti2018AA.609A.10V}}). The IR excesses in individual bands were calculated using the $\rm Q_{VIJA}$ color indices defined by \citet{Damiani2006} and \citet{GuarcelloMDP2009}. These color indices compare the V-I and J-A colors, with `A' being one of the Spitzer bands. Since these indices increase as J-A becomes more red, and since they are independent of extinction, they can be used to separate the extinguished stellar population from stars with intrinsic red colors and to calculate the excess in individual infrared bands. Among the candidate Class~III/TD objects, 288 stars were observed both with CoRoT and $Chandra$. Of these, 74 were detected in X-rays with more than 60 counts (66 Class~III and 8 TD objects). These stars constituted the sample of targets studied in this paper. \par

\section{Simultaneous optical and X-ray variability}
\label{III_var}
 
 \subsection{The classification scheme}
 \label{class_sec}
 
     \begin{table}
    \caption{Correlation tests of the anti-correlated stars. N$\rm_{phot}$ is the number of X-ray counts in each time interval.}
    \label{class_table_ac}
    \centering                       
    \begin{tabular}{ccc}        
    \hline                
    N$\rm_{phot}^1$ & $r$ & P($r$) \\
    \hline\hline
    \multicolumn{3}{l}{Mon-159} \\
    \hline
    20  & -0.61 & 0.06 \\
    50  & -0.90 & 0.04 \\
    \hline                                  
    \multicolumn{3}{l}{Mon-198} \\
    \hline
    20  & -0.18 & 0.07 \\
    30  & -0.25 & 0.04 \\
    40  & -0.34 & 0.02 \\
    50  & -0.42 & 0.01 \\
    60  & -0.40 & 0.02 \\
    80  & -0.58 & 0.00 \\
    100 & -0.72 & 0.00 \\
    \hline                                  
    \multicolumn{3}{l}{Mon-220} \\
    \hline
    20  &-0.53 & 0.06  \\
    30  &-0.73 & 0.01  \\
    40  &-0.70 & 0.01  \\
    50  &-0.72 & 0.03  \\
    \hline                                  
    \multicolumn{3}{l}{Mon-328} \\
    \hline
    20  &-0.86 & 0.00  \\
    30  &-0.82 & 0.00  \\
    40  &-0.71 & 0.00  \\
    50  &-0.94 & 0.00  \\
    60  &-0.71 & 0.01  \\
    \hline              
    \multicolumn{3}{l}{Mon-368} \\
    \hline
    50 &  -0.90 & 0.04 \\
    \hline                                  
    \multicolumn{3}{l}{Mon-389} \\
    \hline
    20 & -0.71 & 0.00  \\
    30 & -0.86 & 0.01  \\
    40 & -0.89 & 0.02  \\
    50 & -0.90 & 0.04  \\
    \hline                                  
    \multicolumn{3}{l}{Mon-548} \\
    \hline
    30 &  -0.60& 0.09  \\
    \hline              
    \multicolumn{3}{l}{Mon-657} \\
    \hline
    20  &-0.40 & 0.10  \\
    60  &-1.00 & 0.00  \\
    \hline              
         
    \end{tabular}
    \begin{tabular}{ccc}        
    \hline 
    N$\rm_{phot}$ & $r$ & P($r$)  \\
    \hline\hline
    
    \multicolumn{3}{l}{Mon-680} \\
    \hline
    20 & -0.60 & 0.01 \\
    40 & -0.75 & 0.05 \\
    \hline              
    \multicolumn{3}{l}{Mon-697} \\
    \hline
    20 & -0.64 & 0.00 \\
    30 & -0.58 & 0.06 \\
    40 & -0.81 & 0.01 \\
    \hline                                  
    \multicolumn{3}{l}{Mon-724} \\
    \hline
    20 & -0.60 & 0.00 \\
    30 & -0.63 & 0.01 \\
    40 & -0.77 & 0.00 \\
    50 & -0.73 & 0.02 \\
    60 & -0.71 & 0.07 \\
    \hline                                  
    \multicolumn{3}{l}{Mon-770} \\
    \hline
    20  &-0.38 & 0.09 \\
    30  &-0.67 & 0.01 \\
    50  &-0.89 & 0.02 \\
    60  &-0.80 & 0.10 \\
    \hline                                  
    \multicolumn{3}{l}{Mon-938} \\
    \hline
    20 & -0.77 & 0.07 \\
    \hline              
    \multicolumn{3}{l}{Mon-1027} \\
    \hline
    20  & -0.54 & 0.09 \\
    \hline
    \multicolumn{3}{l}{Mon-1218} \\
    \hline
    20  &-0.50 & 0.01  \\
    30  &-0.69 & 0.00  \\
    40  &-0.72 & 0.01  \\
    50  &-0.58 & 0.10  \\
    \hline                                  
    \multicolumn{3}{l}{Mon-1359} \\
    \hline
    20  &-0.40 & 0.10  \\
    40  &-0.89 & 0.02  \\
    50  &-0.89 & 0.02  \\
    \hline              
    \end{tabular}
    \end{table}

     \begin{table}
    \caption{Correlation tests of the correlated stars}
    \label{class_table_cor}
    \centering                       
    \begin{tabular}{ccc}        
    \hline                
    N$\rm_{phot}$ & $r$ & P($r$) \\
    \hline\hline
    \multicolumn{3}{l}{Mon-263} \\
    \hline
    20  & 0.49 & 0.05  \\
    30  & 0.63 & 0.04  \\
    100 & 0.40 & 0.60  \\
    \hline                                  
    \multicolumn{3}{l}{Mon-394} \\
    \hline
    20 &  0.83 & 0.04 \\
    \hline                                  
    \multicolumn{3}{l}{Mon-430} \\
    \hline
    20 &  0.55 & 0.04 \\
    \hline                                  
    \multicolumn{3}{l}{Mon-448} \\
    \hline
    20  & 0.46 & 0.00  \\
    30  & 0.68 & 0.00  \\
    40  & 0.76 & 0.00  \\
    50  & 0.83 & 0.00  \\
    60  & 0.82 & 0.00  \\
    80  & 0.83 & 0.00  \\
    100 & 0.84 &  0.00 \\
    \hline              
    \multicolumn{3}{l}{Mon-499} \\
    \hline
    20 &  0.77 & 0.07 \\
    \hline              
    \multicolumn{3}{l}{Mon-749} \\
    \hline
    20 &  0.48 & 0.04 \\
    30 &  0.64 & 0.03 \\
    \hline              
    \multicolumn{3}{l}{Mon-809} \\
    \hline
    20  & 0.40 & 0.00  \\
    30  & 0.48 & 0.00  \\
    40  & 0.58 & 0.00  \\
    50  & 0.61 & 0.00  \\
    60  & 0.66 & 0.00  \\
    80  & 0.71 & 0.00  \\
    100 &  0.68&  0.00 \\
    \hline                                  
    
    \end{tabular}
    \begin{tabular}{ccc}        
    \hline 
    N$\rm_{phot}$ & $r$ & P($r$) \\
    \hline\hline
    
    \multicolumn{3}{l}{Mon-810} \\
    \hline
    20 &  0.55 & 0.00 \\
    30 &  0.62 & 0.00 \\
    40 &  0.67 & 0.00 \\
    50 &  0.68 & 0.00 \\
    60 &  0.65 & 0.00 \\
    80 &  0.76 & 0.00 \\
    100&  0.67 & 0.01 \\
    \hline                                  
    \multicolumn{3}{l}{Mon-869} \\
    \hline
    40 &  0.83 & 0.04 \\
    \hline              
    \multicolumn{3}{l}{Mon-881} \\
    \hline
    20  & 0.64 & 0.00  \\
    30  & 0.62 & 0.03  \\
    60  & 0.90 & 0.04  \\
    80  & 1.00 & 0.00  \\
    \hline                                  
    \multicolumn{3}{l}{Mon-1275} \\
    \hline
    20  & 0.56 & 0.00  \\
    30  & 0.57 & 0.00  \\
    40  & 0.58 & 0.00  \\
    50  & 0.65 & 0.00  \\
    60  & 0.64 & 0.00  \\
    80  & 0.65 & 0.00  \\
    100 &  0.59&  0.02 \\
    \hline
    \end{tabular}
    \end{table}

     \begin{table}
    \caption{Correlation tests of the not correlated stars}
    \label{class_table_nc}
    \centering                       
    \begin{tabular}{ccc}        
    \hline
    N$\rm_{phot}$ & $r$ & P($r$) \\
   \hline\hline                
   \multicolumn{3}{l}{Mon-102} \\  
   \hline
   20 &-0.13 & 0.52 \\
   \hline
   \multicolumn{3}{l}{Mon-108} \\  
   \hline
   20 &-0.18 & 0.70 \\
   \hline
   \multicolumn{3}{l}{Mon-111} \\  
   \hline
   20 &-0.14 & 0.79 \\
   \hline
   \multicolumn{3}{l}{Mon-223} \\  
   \hline
   20 &-0.03 & 0.88 \\
   30 &-0.16 & 0.51 \\
   40 &-0.17 & 0.58 \\
   50 &-0.05 & 0.88 \\
   60 &-0.25 & 0.52 \\
   80 &-0.20 & 0.70 \\
   100&-0.20 & 0.80 \\
   \hline
   \multicolumn{3}{l}{Mon-241} \\  
   \hline
   20 &-0.14 & 0.15 \\
   30 &-0.08 & 0.50 \\
   40 & 0.11 & 0.37 \\
   50 & 0.08 & 0.54 \\
   60 & 0.10 & 0.53 \\
   80 & 0.19 & 0.25 \\
   100& 0.18 & 0.36 \\
   \hline
   \multicolumn{3}{l}{Mon-307} \\
   \hline
   20 & 0.54 & 0.22 \\
   30 & 0.19 & 0.65 \\
   40 & 0.10 & 0.87 \\
   50 & 0.20 & 0.75 \\
   60 & 0.50 & 0.67 \\
   80 &-0.25 & 0.25 \\
   100&-0.32 & 0.20 \\
   \hline
   \multicolumn{3}{l}{Mon-459} \\
   \hline
   20 & 0.50 & 0.67 \\
   30 &-0.30 & 0.62 \\
   40 & 0.20 & 0.75 \\
   50 & 0.09 & 0.87 \\
   60 & 0.10 & 0.87 \\
   \hline
   \multicolumn{3}{l}{Mon-477} \\
   \hline
   20 &-0.18 & 0.08 \\
   30 &-0.21 & 0.09 \\
   40 &-0.20 & 0.18 \\
   50 &-0.23 & 0.18 \\
   60 &-0.27 & 0.15 \\
   80 &-0.25 & 0.25 \\
   100&-0.32 & 0.20 \\
   \hline
   \multicolumn{3}{l}{Mon-517} \\
   \hline
   20 & 0.16 & 0.36 \\
   30 & 0.05 & 0.81 \\
   40 & 0.03 & 0.92 \\
   50 & 0.20 & 0.53 \\
   60 & 0.33 & 0.42 \\
   80 &-0.04 & 0.94 \\
   100& 0.80 & 0.20 \\
   \hline
    
    \end{tabular}
    \begin{tabular}{ccc}        
    \hline 
    N$\rm_{phot}$ & $r$ & P($r$) \\
    \hline\hline
    
   \multicolumn{3}{l}{Mon-519} \\
   \hline
   20 &-0.14 & 0.69 \\
   30 &-0.09 & 0.87 \\
   40 & 0.50 & 0.67 \\
   \hline
   \multicolumn{3}{l}{Mon-606} \\
   \hline
   20 & 0.11 & 0.42 \\
   30 &-0.07 & 0.61 \\
   40 &-0.13 & 0.40 \\
   50 &-0.11 & 0.54 \\
   60 &-0.22 & 0.26 \\
   80 &-0.11 & 0.64 \\
   100&-0.16 & 0.55 \\
   \hline
   \multicolumn{3}{l}{Mon-695} \\
   \hline
   20 & 0.23 & 0.16 \\
   30 & 0.30 & 0.11 \\
   40 & 0.24 & 0.29 \\
   50 & 0.36 & 0.16 \\
   60 & 0.40 & 0.15 \\
   80 & 0.20 & 0.61 \\
   100& 0.36 & 0.43 \\
   \hline
   \multicolumn{3}{l}{Mon-745} \\
   \hline
   20 & 0.15 & 0.43 \\
   30 & 0.19 & 0.43 \\
   40 & 0.25 & 0.38 \\
   50 & 0.12 & 0.77 \\
   \hline
   \multicolumn{3}{l}{Mon-819} \\
   \hline
   20 &-0.08 & 0.39 \\
   30 &-0.11 & 0.34 \\
   40 &-0.08 & 0.56 \\
   50 &-0.09 & 0.56 \\
   60 &-0.13 & 0.46 \\
   80 &-0.10 & 0.61 \\
   100&-0.03 & 0.91 \\
   \hline
   \multicolumn{3}{l}{Mon-919} \\
   \hline
   20 &-0.04 & 0.89 \\
   30 &-0.07 & 0.87 \\
   40 &-0.09 & 0.87 \\
   50 &-0.10 & 0.87 \\
   60 & 0.10 & 0.87 \\
   \hline
   \multicolumn{3}{l}{Mon-926} \\
   \hline
   20 & 0.18 & 0.64 \\
   30 & 0.40 & 0.60 \\
   \hline
   \multicolumn{3}{l}{Mon-1279} \\
   \hline
   20 & 0.02 & 0.87 \\
   30 & 0.03 & 0.89 \\
   40 & 0.07 & 0.74 \\
   50 & 0.41 & 0.17 \\
   60 &-0.09 & 0.75 \\
   80 & 0.06 & 0.85 \\
   100& 0.28 & 0.46 \\
   \hline    
   \end{tabular}
   \end{table}
   
    In this section we describe how we selected sources whose simultaneous optical and X-ray variability is strongly correlated, strongly anti-correlated, or does not present any evident correlation. We classified these stars in three groups: the ``anti-correlated'', ``correlated'', and ``not correlated'' stars, respectively. To this aim, we adopted an approach similar to that of \citet{Guarcello2017AA.602A.10G}. We first divided the CoRoT and $Chandra$ light curves into suitable time intervals. The number of adopted intervals is set by fixing the number (N$\rm_{phot}$) of X-ray photons detected during each time interval in the broad energy band (0.5-7.9$\,$keV). For each star, we tested up to seven values of N$\rm_{phot}$: 20, 30, 40, 50, 60, 80, and 100 counts\footnote{Also larger values of N$\rm_{phot}$ were tested during the analysis but they are not discussed here since they were never used in the final classification scheme}. For each value of N$\rm_{phot}$, we calculated and compared in each time interval the median CoRoT ``whiteflux'' and the average X-ray photon flux. We then investigated any existing correlation between the optical and X-ray variability, performing a two-sided Spearman's rank correlation test.\par
    
    We removed from the correlation tests the time intervals where flares were detected. X-ray flares were automatically detected using the approach defined in \citet{CaramazzaFMR2007AA}, that is, by dividing the X-ray light curve into blocks statistically compatible with having a constant light curve (maximum likelihood blocks) and classifying these intervals according to the measured count-rates and their time derivative (a detailed study of the flares observed in NGC~2264 is presented in \citealp{Flaccomio2018AA.620A.55F}). Optical flares were selected via visual inspection of the light curves. \par
   
    We first selected the stars for which all tests yield P($r$)$\leq$0.1 as those with a strong positive or negative correlation between the optical and X-ray variability. Among these stars, we sorted  in the anti-correlated group those for which all the tested N$\rm_{phot}$ resulted in correlation coefficients r$<$-0.5, and in the correlated group those for which all tests resulted in correlation coefficients r$>$0.5. These limits were chosen by inspection of the resulting selections. Tables \ref{class_table_ac} and \ref{class_table_cor} show the tests\footnote{As shown in these tables, more restrictive thresholds could have been chosen losing only a small number of stars. The adopted threshold was dictated also by the small number of stars falling in each group.} for the 16 anti-correlated stars and the 11 correlated stars, respectively. We selected the stars with uncorrelated optical and X-ray variability as those fulfilling both the following requirements: 1) no test resulted in r$>$0.5 or r$<$-0.5 and P($r$)$\leq$0.1; 2) in most of the tests, the correlation coefficient being in the -0.3$<$r$<$0.3 range. Adopting these criteria, 16 stars were included in the not correlated group. Their tests are listed in Table \ref{class_table_nc}. The 30 sources not satisfying these requirements are likely stars with a moderated correlation between optical and X-ray variability. However, we did not classify these stars in any of the three groups, and we refer to these stars as not classified. For a few stars, we applied ad-hoc classification rules. For four stars (Mon-198, Mon-657, Mon-770, and Mon-1359) some of the tests fulfilled the requirements of not correlated or not classified stars, while others met the requirements for the anti-correlated stars. We decided to classify there stars in the latter group after a visual inspection of their variability. A similar situation occurred for two stars (Mon-263, and Mon-749) but in the correlated group. Two stars (Mon-448 and Mon-809) were classified in the correlated sample even if one of seven tests (for Mon-448) and two of seven tests (for Mon-809) fulfilled the requirements for the not correlated stars. For the star Mon-777, we did not consider the 2008 $Chandra$ observations, which were dominated by a stellar flare and a CoRoT flux discontinuity. Among the stars with transition disks, two were classified in the anti-correlated sample (Mon-328 and Mon-697), one in the correlated (Mon-448), and two in the ``not-correlated'' stars (Mon-919 and Mon-926). \par
    
    \begin{figure}[]
        \centering      
        \includegraphics[width=9cm]{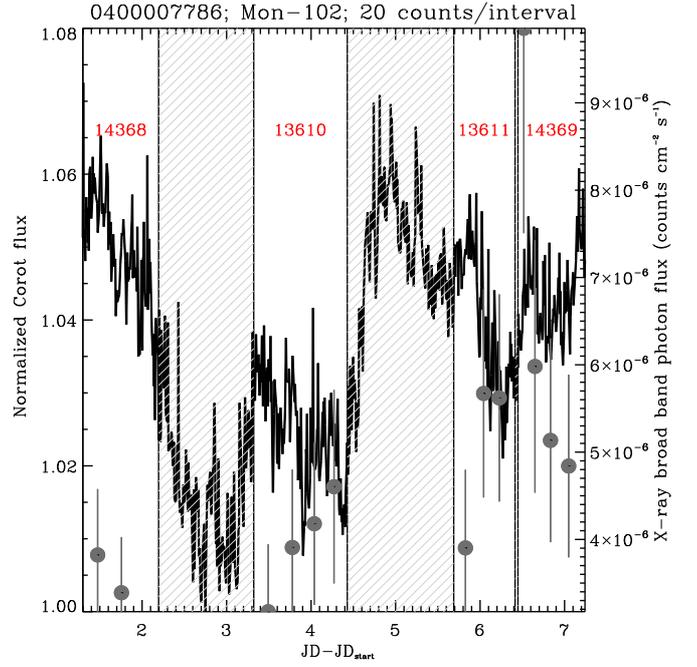}
        \caption{Optical and X-ray light curve of the star Mon-102. The black line shows the CoRoT light curve. The time windows corresponding to the $Chandra$ observations are the unshaded intervals. The gray dots mark the observed average X-ray photon fluxes in the broadband in the time intervals defined by setting the number of photons per interval to N$\rm_{phot}$=20 (circled if an X-ray flare occurred during the corresponding time interval; see the figures in Appendix \ref{appendixA}). The red numbers in the top indicate the Chandra Obs.ID of the given Chandra observation.}
        \label{lc_mon102}
        \end{figure}

        In order to associate an error with the correlation coefficient, we repeated each correlation test 5000 times, replacing each flux value (both in optical and X-rays) with a value obtained randomly from a normal distribution centered on the nominal flux value and with a width equal to the flux error. The error associated with the correlation coefficient is then calculated as the standard deviation of the distribution obtained from 5000 simulated correlation coefficients. \par
    
    As an example, Fig. \ref{lc_mon102} shows the CoRoT light curve of the star in our sample with the smallest Mon- number, classified as a not correlated source. In Appendix \ref{appendixA}, we show all light curves observed during the $Chandra$ observations, while in Appendix \ref{appendixC}, we show their entire $\sim$22-day  CoRoT light curves. \par

        \begin{figure*}[]
        \centering      
        \includegraphics[width=18cm]{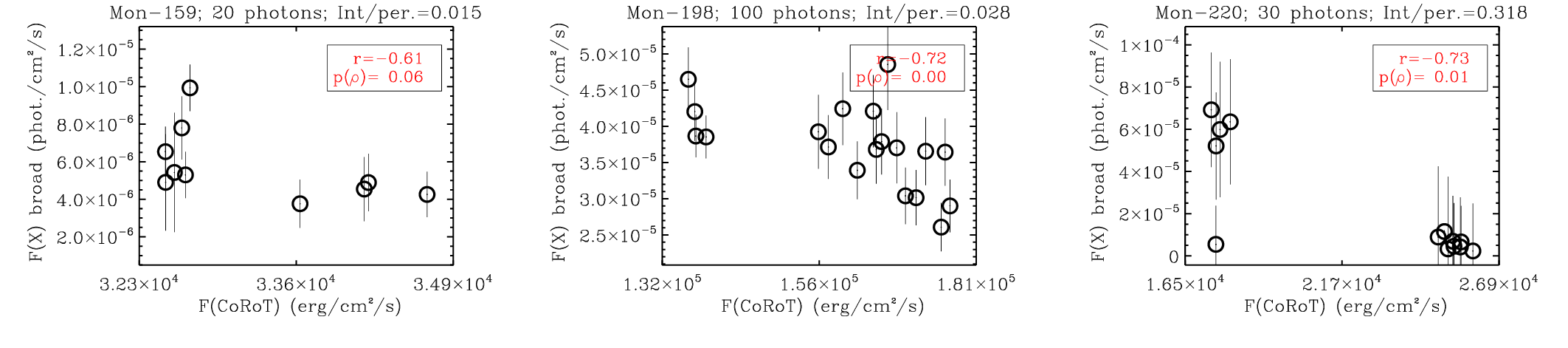}
        \includegraphics[width=18cm]{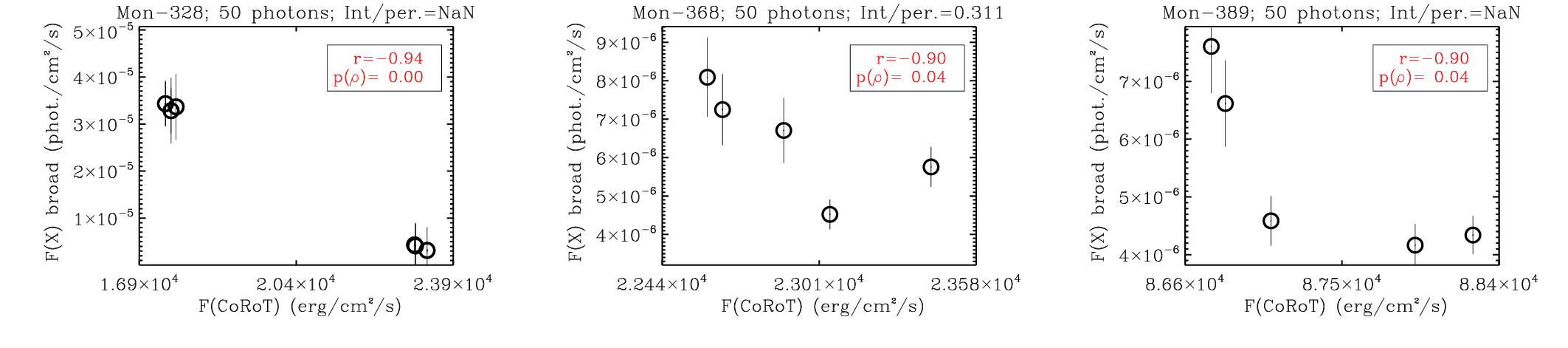}        
        \includegraphics[width=18cm]{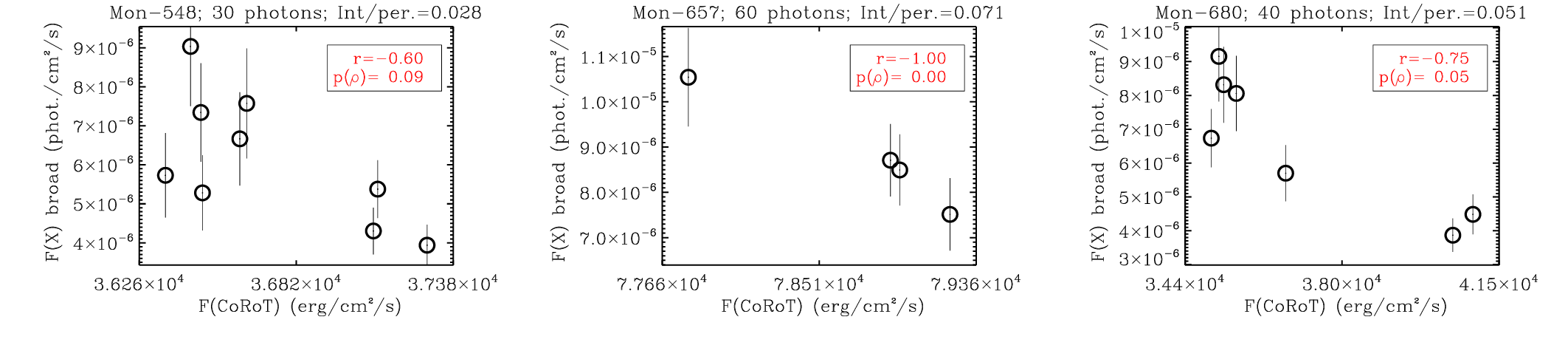}        
        \includegraphics[width=18cm]{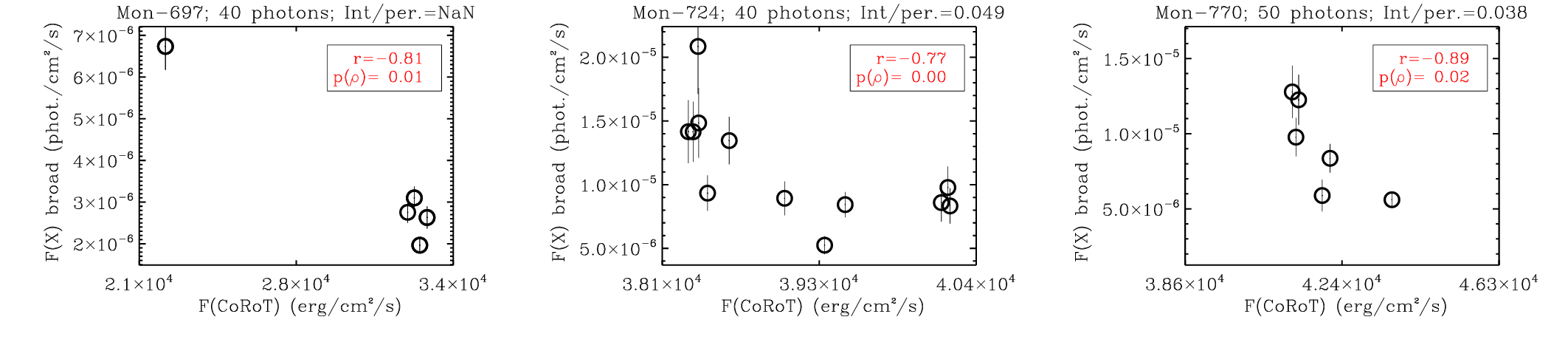}        
        \includegraphics[width=18cm]{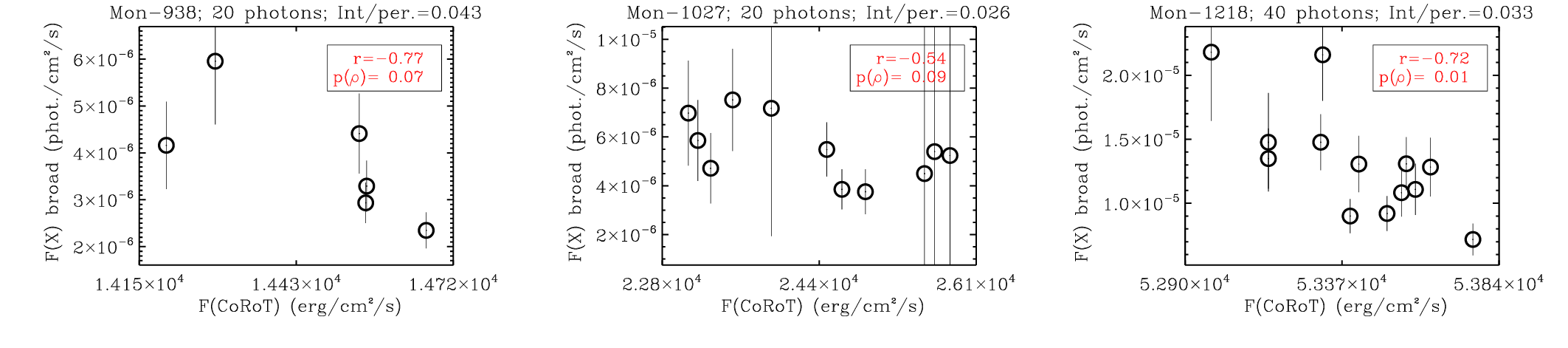}
        \includegraphics[width=18cm]{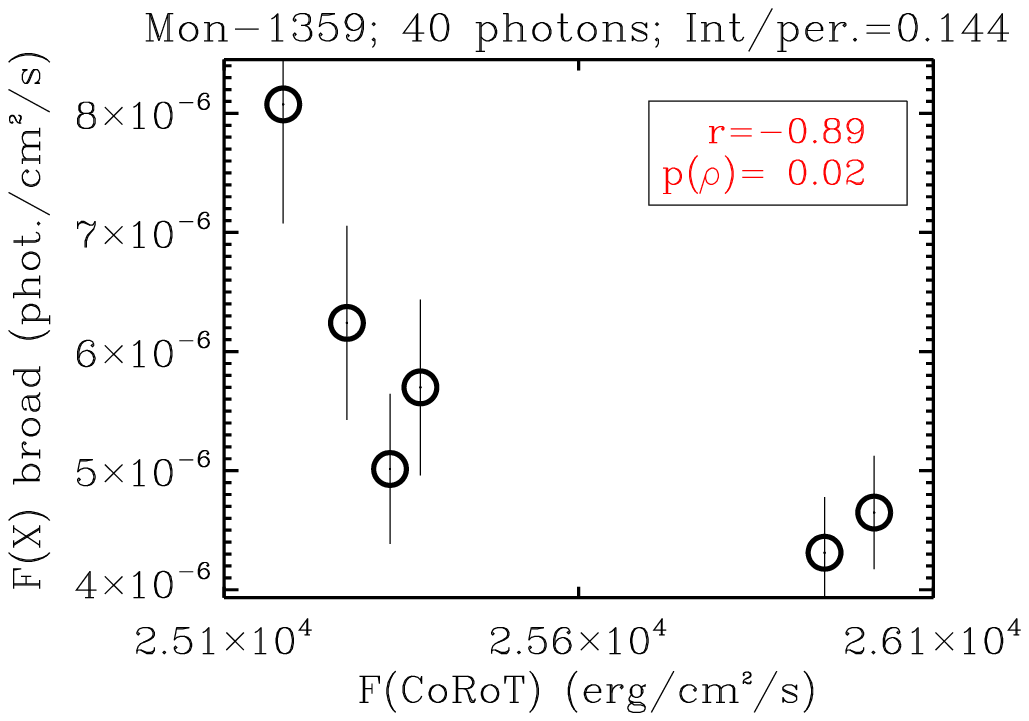}
        \caption{Optical (x axis) and X-ray (y axis) average fluxes observed in the defined time intervals for the stars with anti-correlated X-ray and optical variability. The  Mon-name of the star, the number of photons contained in each time interval (N$\rm_{phot}$), the median duration of the time intervals in units of the rotation period, and the results of the Spearman correlation test are also shown.}
        \label{variab_cl3anticorr1}
        \end{figure*}

%       \begin{figure*}[]
%       \centering      
%       \includegraphics[width=18cm]{lightcurve_correlations_anticorrelated6.ps}        
%       \caption{Optical vs. X-ray median fluxes observed in the defined time intervals for the stars with anti-correlated X-ray and optical variability. Panel format and content generally follows Fig.~\ref{variab_cl3anticorr1}.}
%       \label{variab_cl3anticorr2}
%       \end{figure*}
  
        \begin{figure*}[]
        \centering      
        \includegraphics[width=18cm]{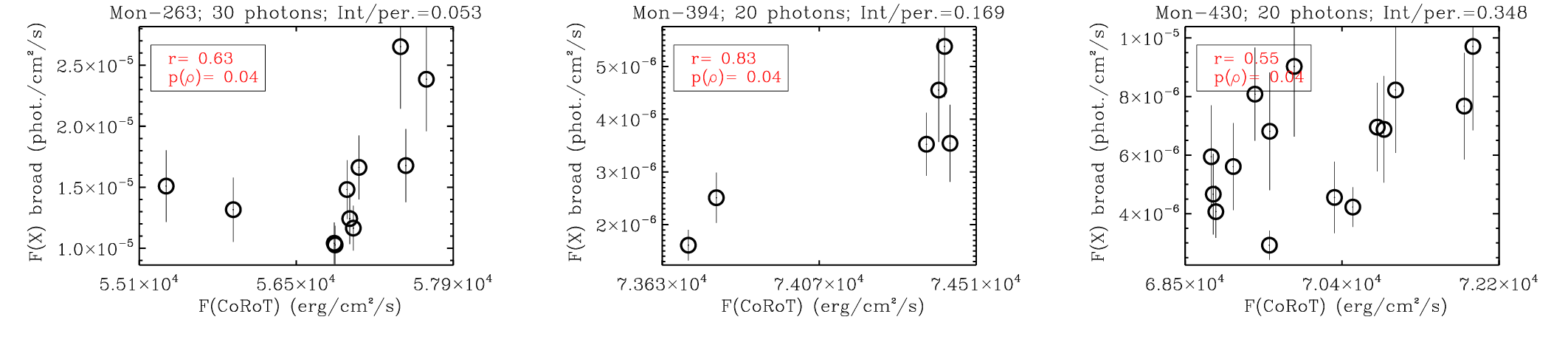}
    \includegraphics[width=18cm]{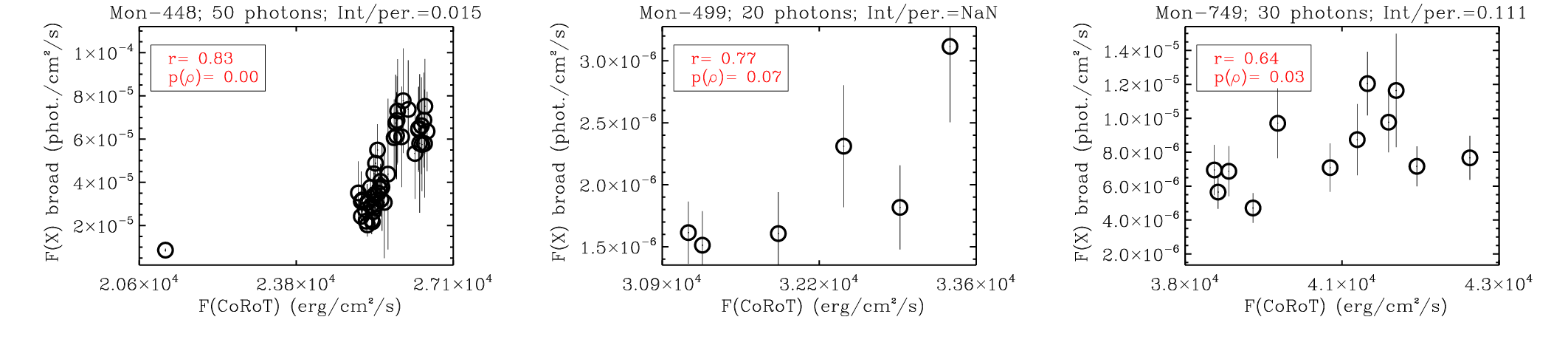}
        \includegraphics[width=18cm]{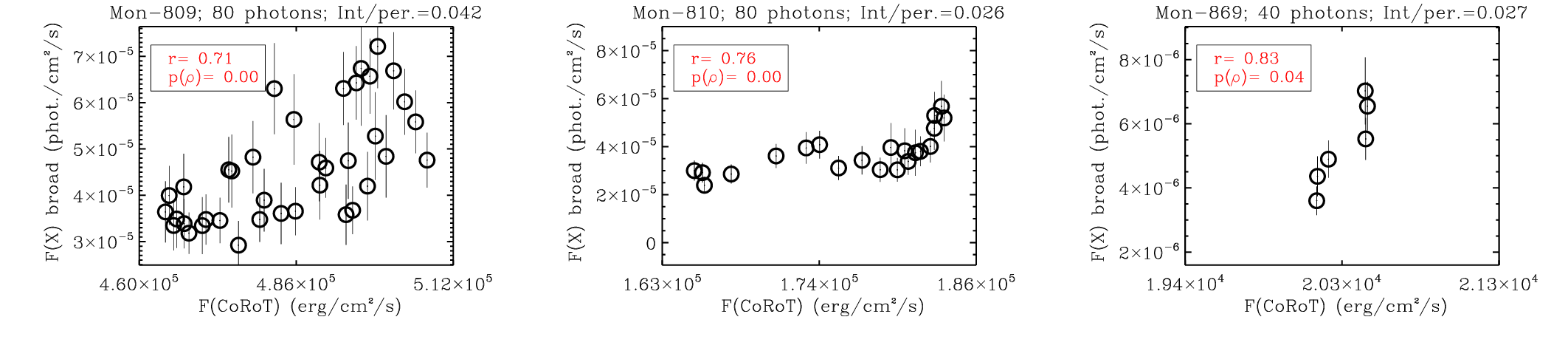}
        \includegraphics[width=18cm]{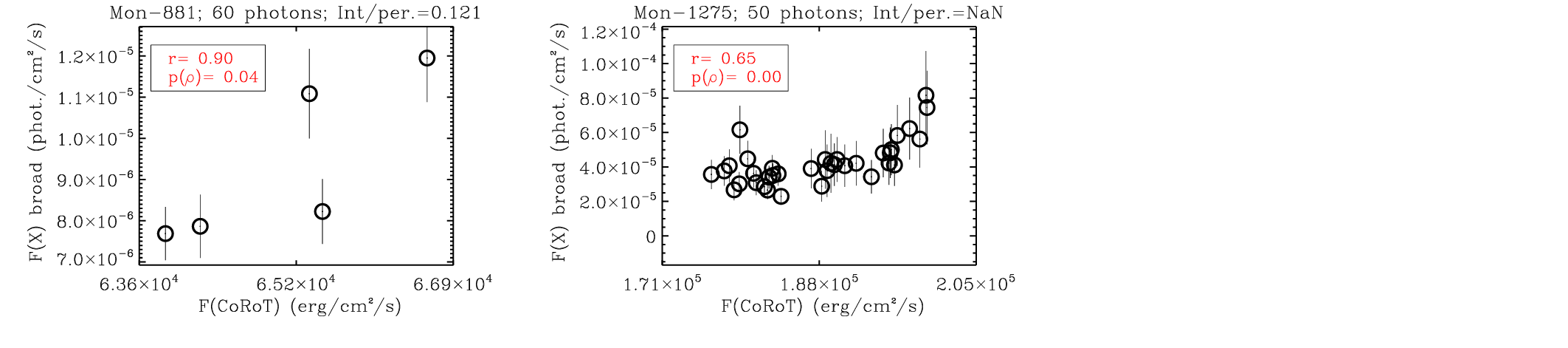}
        \caption{Optical (x axis) and X-ray (y axis) average fluxes observed in the defined time intervals for the stars with correlated X-ray and optical variability. The  Mon-name of the star, the number of photons contained in each time interval (N$\rm_{phot}$), and the results of the Spearman correlation test are also shown.}
        \label{variab_cl3corr}
        \end{figure*}

\begin{figure*}[]
    \centering
    \includegraphics[width=18cm]{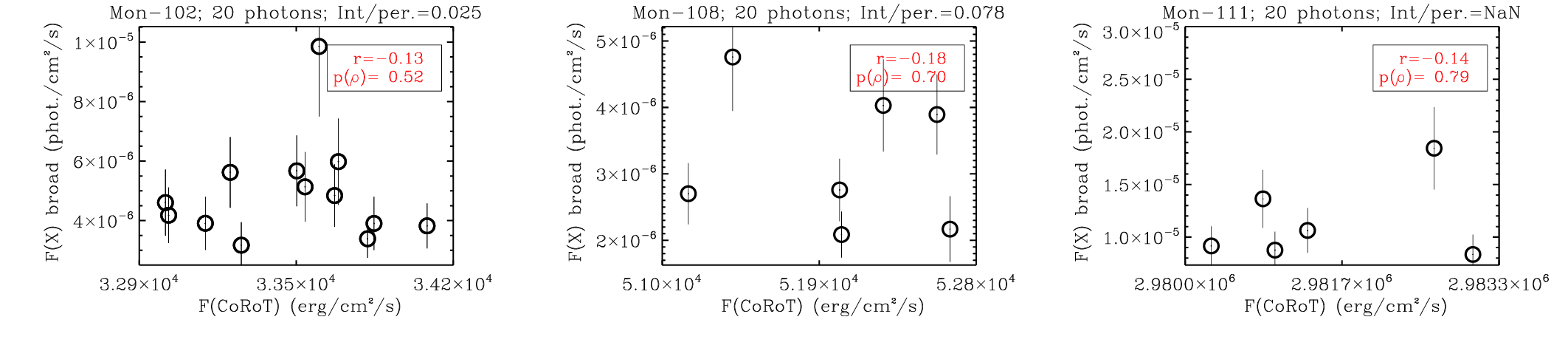}
    \includegraphics[width=18cm]{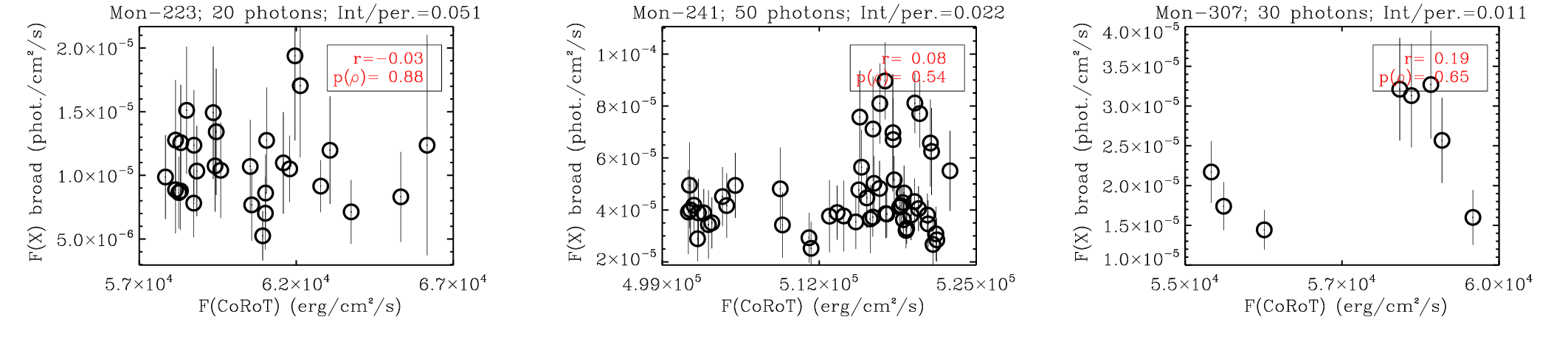}
    \includegraphics[width=18cm]{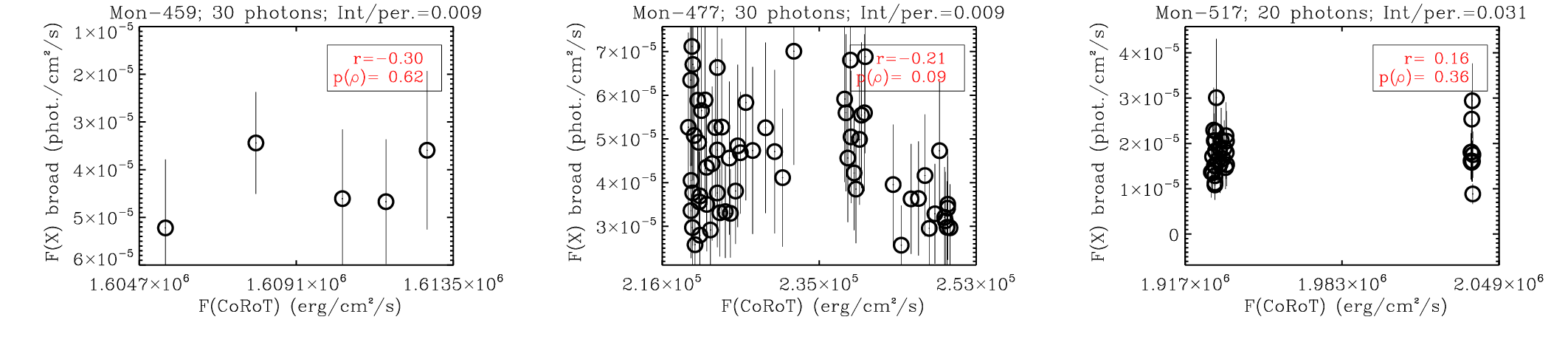}
    \includegraphics[width=18cm]{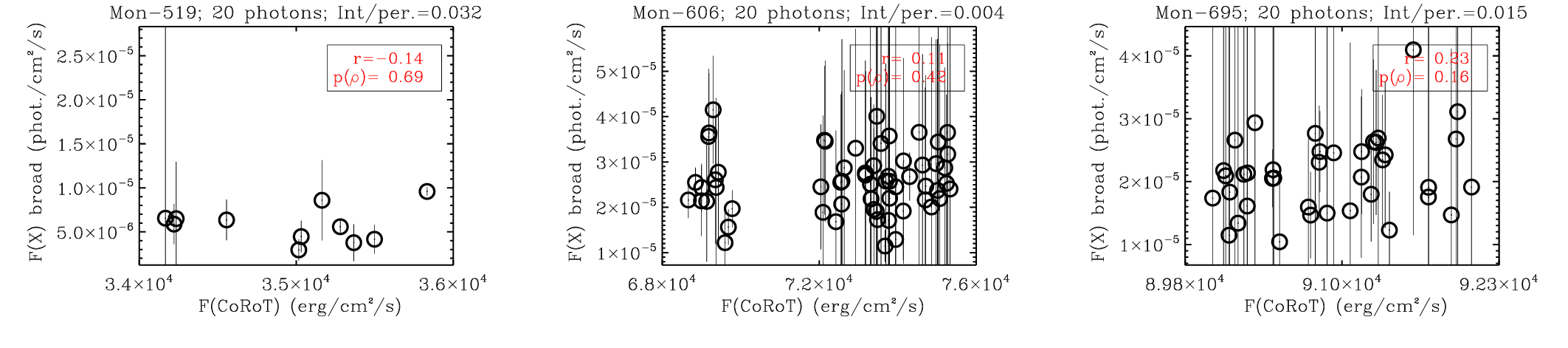}
    \includegraphics[width=18cm]{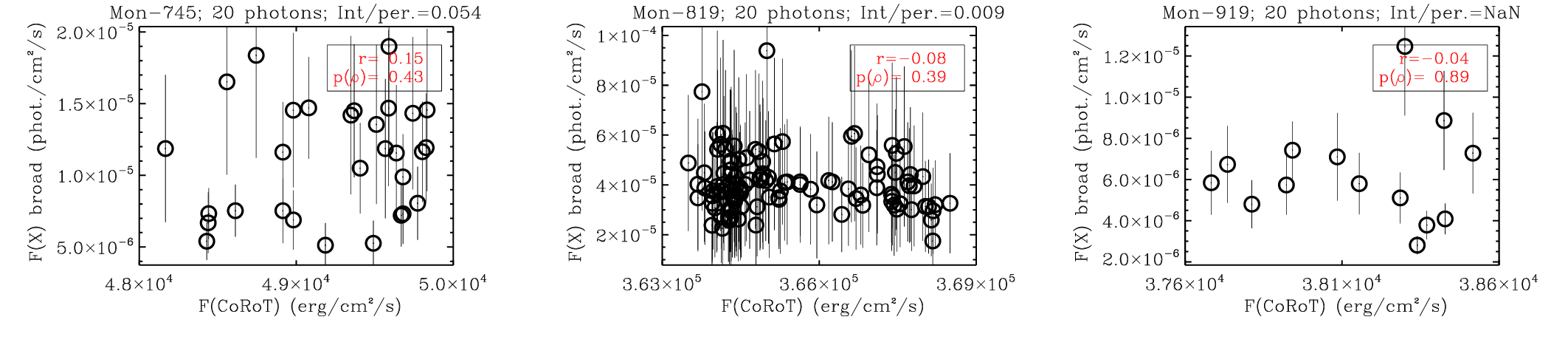}
    \includegraphics[width=18cm]{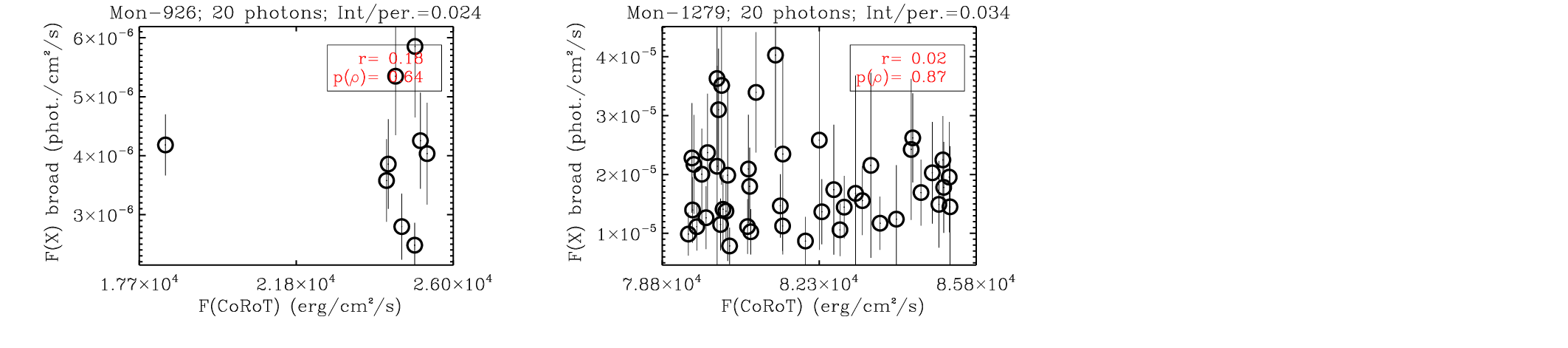}
    \caption{Optical (x axis) and X-ray (y axis) average fluxes observed in the defined time intervals for the stars in the not correlated group. The Mon-name of the star, the number of photons contained in each time interval (N$\rm_{phot}$), and the results of the Spearman correlation test are also shown.}
    \label{variab_cl3noncorr1}
    \end{figure*}

The optical and X-ray fluxes observed during the time intervals defined to sample the light curves of the 16 anti-correlated stars are shown in Fig. \ref{variab_cl3anticorr1}, those of the 11 correlated stars in Fig \ref{variab_cl3corr}, and those of the 17 not correlated stars in Fig. \ref{variab_cl3noncorr1}. In these figures, each panel corresponds to one star, whose Mon- name is shown in the title, with black circles marking the median CoRoT flux and the average X-ray photon flux observed in each time interval. Error bars are large enough to be visible only in X-rays. The obtained Spearman's correlation coefficient and its significance are shown in the upper right corner of each panel. \par

%     \begin{figure*}[]
%     \centering
%     \caption{Optical vs. X-ray median fluxes observed in the defined time intervals for the stars with not correlated X-ray and optical variability. Panel format and content generally follows Fig.~\ref{variab_cl3noncorr1}.}
%     \label{variab_cl3noncorr2}
%     \end{figure*}
  
 \subsection{Reliability of the classification scheme}
 \label{simul_sec}

\begin{figure}[h]
    \begin{centering}
    \includegraphics[width=0.5\textwidth]{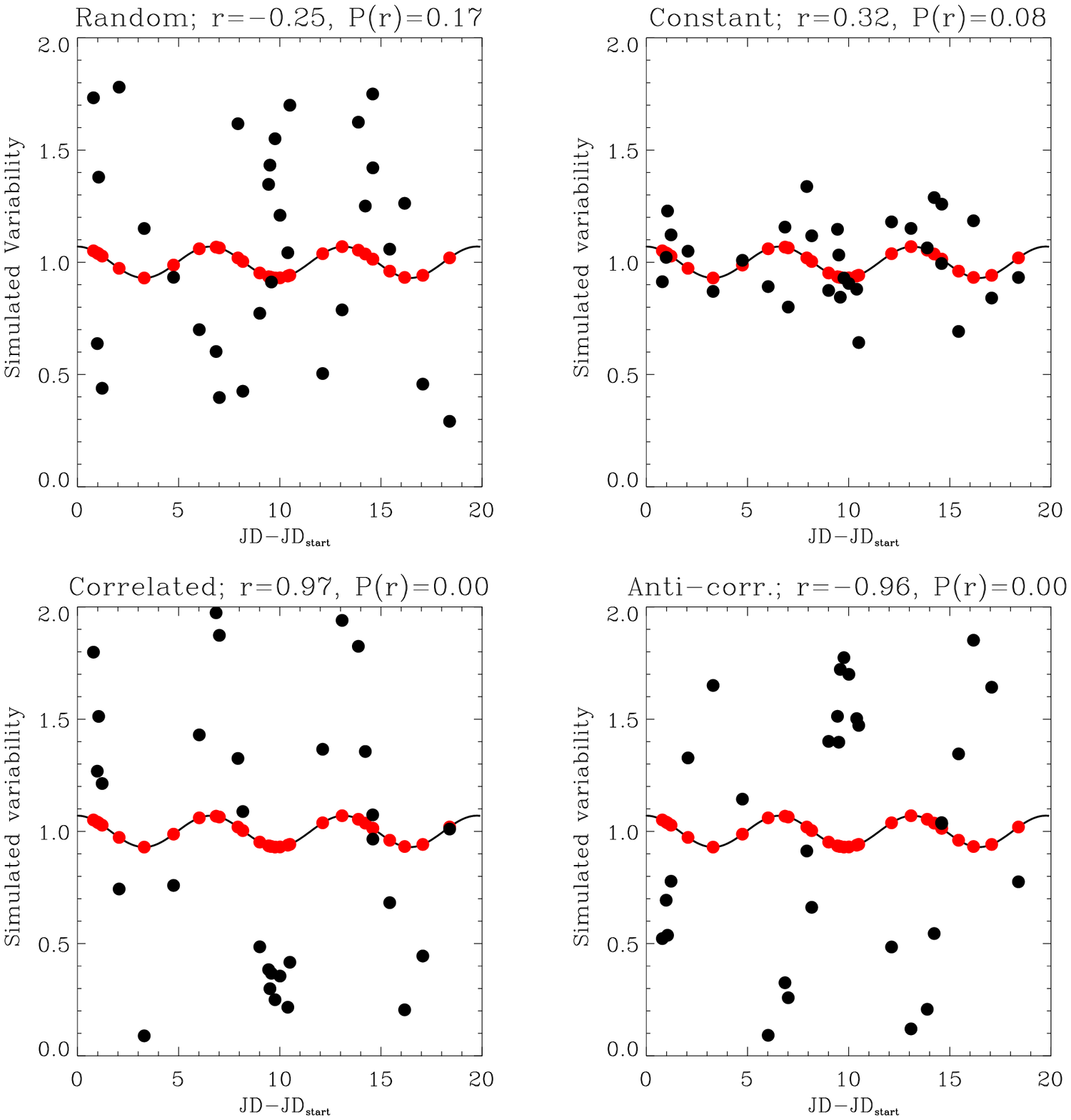}  
    \caption{Correlation tests between a sinusoid with three cycles in 20 days and a random (upper left panel), constant (upper right panel), correlated (bottom left panel), and anti-correlated (bottom right panel) simulated X-ray light curve. The red points show how the sinusoid is sampled, and the black points mark the simulated X-ray photon fluxes. The results of the Spearman two-ranks correlation test are shown above each panel.}
        \label{sim3period_fig1}
        \end{centering}
        \end{figure}

We performed several tests aimed at verifying the reliability of the adopted classification scheme, and quantifying the level of contamination expected in the three groups (e.g., stars whose correlation between optical and X-ray variability was classified incorrectly). The first set of simulations was aimed at estimating the chances of misclassifying the correlation between the optical and X-ray variability in stars with different rotation periods. We modeled the optical variability with sinusoidal functions, with 7\% amplitude (which is equal to the median optical variability amplitude of the stars in the three groups; see Sect. \ref{amplitude_sect}) centered on one, and increasing period between 2 and 20 days. We then sampled the sinusoids using 30 intervals defined randomly (random length, not overlapping). In each interval, we calculated the average value of the sinusoid. We then simulated the X-ray photon fluxes adopting four assumptions: i) random X-ray variability, ii) constant X-ray light curve, iii) correlated, and iv) anti-correlated optical and X-ray variability. In the first case, the X-ray light curve is simulated with random numbers drawn from a uniform distribution centered on one and going from 0.2 to 1.8 (these values are set in order to have an amplitude variability equal to the median X-ray variability observed in these stars). The constant X-ray light curve is simulated setting the X-ray value equal to one, and adding to each point a noise simulated adopting a normal distribution with a width equal to the 95\% quantile of the distribution of the normalized error bars of all the X-ray fluxes measured in the time intervals. The correlated and the anti-correlated X-ray light curves were simulated generating random numbers from the same sinusoid used to simulate the optical light curve, but with a phase shift equal to 0 or $\pi/2$, respectively, and adding Gaussian noise as for the constant light curve. Figure \ref{sim3period_fig1} shows one of these simulations performed assuming a sinusoid with a period of 3 days. 

For each sinusoid and for each assumed X-ray correlation (i.e., constant, random, correlated, or anti-correlated), we simulated 5000 X-ray light curves and studied how the resulting Spearman correlation coefficients were distributed. Figures from \ref{appC_fig2} to \ref{appC_fig11} in Appendix \ref{appendixD} show the distributions of the Spearman correlation test coefficients obtained with sinusoids with periods from 1 to 20 days, assuming random, constant, correlated, and anti-correlated simulated X-ray light curves. It is evident that the chances for incorrect classifications are negligible in all the simulated types of variability. \par

 The second set of simulations was aimed at estimating the chances that uncorrelated random optical and X-ray variability could be classified as correlated or anti-correlated variability in our classification scheme, as a function of the X-ray luminosity and the number of intervals used to sample the light curves. To simulate the random optical variability, we generated 10000 random numbers from a normal distribution centered on zero and with a width set equal to the median optical variability amplitude of these stars. The X-ray light curve is simulated assuming three values of total counts: 232, 388, 804, which are the 25\% quantile, median, and 75\% quantile of the distribution of X-ray net counts in our sample of stars. We associated with each simulated X-ray count a ``detection time'' randomly generated between 0 and 10000 from a uniform distribution. We then assumed one of the values of N$\rm_{phot}$, and defined the intervals used to sample the two curves. We calculated for each interval the median flux of the simulated optical curve and the simulated X-ray photon fluxes, normalizing the distribution of the resulting fluxes between -0.4 and 0.4 for simplicity. Once we generated the two simulated sampled light curves, we analyzed their correlation with the Spearman two-ranks correlation test. \par
 
    \begin{figure}[h]
    \includegraphics[width=0.4\textwidth]{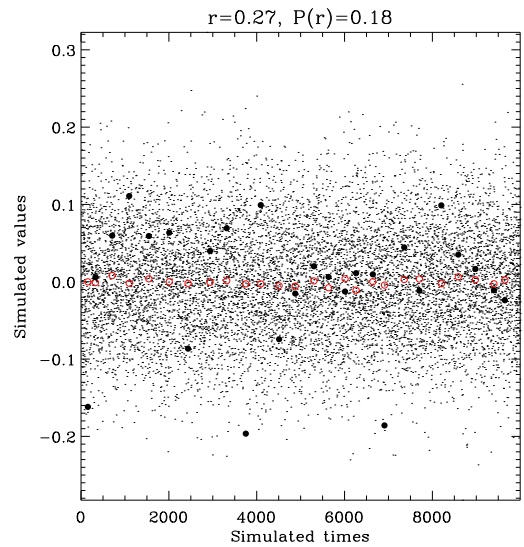}  
    \caption{An example of a simulated random light curve. The small points represent the 10000 random numbers, normally distributed around zero, used to simulated the random optical variability; the empty red circles mark the sampled curve obtained from these numbers; the filled circles marked the sampled curve obtained from 340 points randomly distributed between 0 and 10000. The results of the Spearman correlation test are shown above the panel.}
        \label{sim_rand_fig}
        \end{figure}
 
Figure \ref{sim_rand_fig} shows one of the simulated pairs of random curves. In this case, the correlation test indicates that the two curves are not correlated, but this is not always the case, despite the fact that the two curves were randomly generated. In order to estimate the chances that the correlation test falsely indicates correlated or anti-correlated variability, we simulated 1000 pairs of curves for each combination of X-ray total counts (232, 388, and 804) and N$\rm_{phot}$ (20, 30, 40, 50, 60, 80, and 100). The upper left panel of Fig. \ref{sim_dist1_fig} in Appendix \ref{appendixD} shows the distribution of the Spearman two-ranks correlation coefficient obtained from all the simulations we performed, with a resulting level of contamination of the correlated and anti-correlated samples equal to about 7.9\%, adopting our classification scheme. The other panels show the distributions obtained adopting different values of simulated X-ray counts, independently from the value of N$\rm_{phot}$. Going from the lowest to the highest X-ray total counts, the level of expected contamination of the correlated and anti-correlated groups ranges between 12.7\% and 6.3\% (considering only the tests matching the requirement P$\rm (r)\leq$0.1). In Fig. \ref{sim_dist2_fig} in Appendix \ref{appendixD}, we show the distribution of the correlation coefficient obtained adopting different values of N$\rm_{phot}$, independently from the number of counts in the simulated X-ray curve. The fraction of tests resulting in correlated or anti-correlated variability varies stochastically as a function of N$\rm_{phot}$, going from 11\% for N$\rm_{phot}$=20 to 6.8\% for N$\rm_{phot}$=50. In conclusion, these simulations indicate that the fraction of uncorrelated optical and X-ray light curves that can be sorted as correlated or anti-correlated, and thus the expected level of contamination of these two groups, is in general lower than 10\%, with the exception of the faintest X-ray sources (however there are two stars fainter than 232 net counts in both the anti-correlated and correlated groups, and three in the not correlated group) for which it is about 13\%.

\section{Properties of the stars in the three correlation groups}
\label{propert_sect}

    In this section we describe some properties of the sources classified in the three correlation groups. The complete list of their parameters is shown in Table \ref{starprop_table}.

    \subsection{X-ray activity}
    \label{explan_sect}

\begin{figure}[]
        \centering      
        \includegraphics[width=8cm]{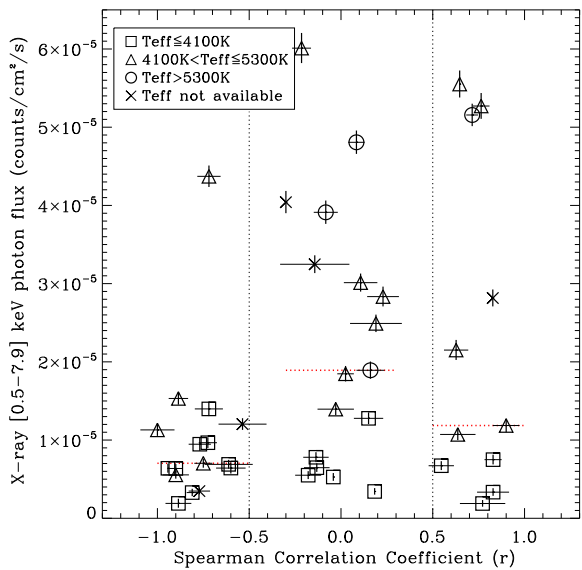}
        \caption{X-ray photon flux vs. the Spearman's correlation coefficient. Different symbols are used to mark stars in three different ranges of T$\rm_{eff}$, as shown in the label. The vertical lines delimit the locus of the anti-correlated and correlated groups. The horizontal dotted lines instead mark the median values of photon flux in the three groups. not correlated stars have on average larger X-ray luminosity than the stars in the other two samples.}
        \label{fxvsr}
        \end{figure}

        The comparison between the X-ray emission among the stars in the three correlation groups can help to elucidate whether the different correlation between optical and X-ray variability is related to different levels of X-ray emission. We thus show in Fig. \ref{fxvsr} the broadband X-ray photon flux of the stars in the three correlation groups, with the dotted horizontal lines marking the median photon fluxes for each group. Sources in the not correlated group show a larger median value of observed photon fluxes than the stars in the other two groups, indicating that these stars are on average brighter in X-rays and thus are characterized by more intense magnetic activity. It is important to note that since in the three groups the amount of stars fainter than the 25\% quantile of the overall X-ray net counts distribution is similar\footnote{The total net counts used in Sect. \ref{simul_sec} are calculated over the intervals used to define the correlation, while the X-ray photon fluxes discussed in Sect. \ref{explan_sect} are calculated over the entire $Chandra$ observations} (Sect. \ref{simul_sec}), it is unlikely that the three groups suffer different incompleteness in the faint end of the X-ray luminosity distribution. A Kolmogorov-Smirnov test on the distribution of the photon fluxes of the not correlated sources versus those of the other two groups together marginally confirms that the two distributions are not drawn from the same parent population (the resulting KS statistic is 0.41, with a significance level of 4\%). As a further test, in Fig. \ref{fxvsr} we plot stars in three different temperature ranges with different
symbols: T$\rm_{eff}\leq$4100$\,$K, 4100$\,$K$<$T$\rm_{eff}\leq$5300$\,$, and T$\rm_{eff}$$>$5300$\,$K, roughly corresponding to M, K, and G stars, respectively. Effective temperatures were obtained and published from the fourth internal data release of the Gaia-ESO Survey (GES) campaign \citep{GilmoreRAB2012} by \citet{Venuti2018AA.609A.10V}.     The larger X-ray emission of the not correlated sources may arise from a different distribution in spectral types of the stars in the three correlation groups. Considering only the M stars,  no difference is evident. Instead, the not correlated group contains a larger number of K and G stars than the other two groups (respectively, 9 GK stars and 6 M stars are in the not correlated sample, 6 GK and 4 M in the correlated group, and 5 GK and 9 M in the anti-correlated group), plus two stars with no temperature available but with an X-ray photon flux typical of the GK stars.
                
% \clearpage

\begin{figure}[]
        \centering      
        \includegraphics[width=8cm]{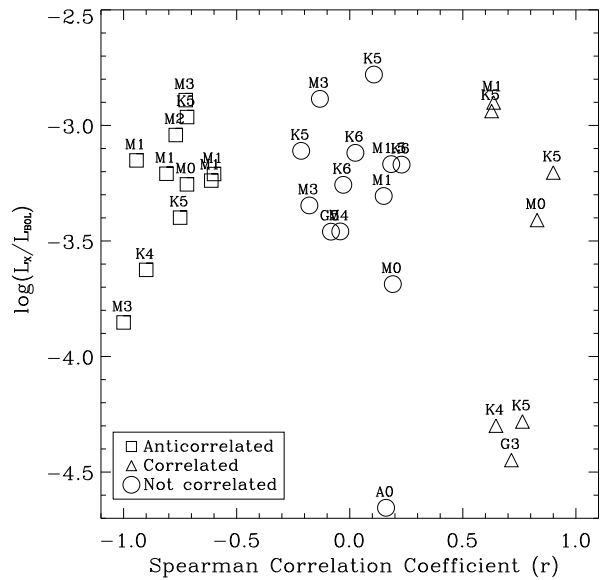}
        \caption{X-ray over bolometric luminosity ratio vs. Spearman's correlation coefficient for the stars in the three correlation groups, marked with squares (anti-correlated), triangles (correlated) and circles (not correlated). The spectral type of each star is also indicated. Stars are in the saturated regime independently from the observed correlation. The A0 star is likely a binary.}
        \label{lxlbolvsr}
        \end{figure}

    We also analyzed the log$\rm\left(L_X/L_{bol} \right)$, which, as shown in Fig. \ref{lxlbolvsr}, is typical of a saturated young stellar population, with log$\rm\left(L_X/L_{bol} \right) \approx -3$ \citep{PizzolatMMS2003AA}, independently of the type of correlation observed between optical and X-ray variability. Stellar X-ray luminosities were obtained using Xspec v.12.8.1. \citep{Arnaud1996} by fitting the observed X-ray spectra averaged over the whole $Chandra$ observations with 1T or 2T APEC thermal plasma model and adopting the WABS model to account for interstellar absorption. More details on the X-ray spectral fitting can be found in \citet{Guarcello2017AA.602A.10G}. Stellar bolometric luminosities were obtained applying a bolometric correction to the available $V$ and $I$ photometry \citep{SungBCK2008}. In both cases, to convert fluxes into luminosity we adopted a distance to NGC~2264 of $760\,$pc, the individual extinctions obtained from known spectral types and photometry \citep{Walker1956ApJS.2.365W,MakidonRSA2004,DahmSimon2005,Venuti2018AA.609A.10V}, and adopting the reddening law of \citet{MunariCarraro1996}. Only a few stars have log$\rm\left(L_X/L_{bol} \right)$ lower than -3.5, which in some cases may be due to binarity. Thus, different correlations between optical and X-ray variability are not related to the saturation of X-ray emission. \par

% \clearpage

\subsection{Stellar mass}
\label{mass_sect}

        In principle, the higher median X-ray luminosity observed in the not correlated stars could be due to the fact that they are on average more massive than those in the other two groups. In fact, since PMS stars are typically in the saturated X-ray regime, their X-ray luminosity does not scale with the rotation period but with stellar mass \citep{PreibischKFF2005}. In addition, stars with different masses may be characterized by different internal structure and this may affect the magnetic field strength and the magnetic activity. We thus estimated the masses of the stars in the three groups by interpolating the optical $V_{0}$ magnitudes and $(V-I)_{0}$ colors with the MIST isochrones \citep{Choi2016ApJ,Dotter2016ApJS.222.8D} with age in the 0.5-15$\,$Myr range. Dereddened magnitudes were calculated from the $V$ and $I$ photometry provided by \citet{SungBCK2008}, adopting the common distance of 760$\,$pc \citep{Sung1997AJ.114.2644S}, and the individual extinctions. In order to estimate the uncertainty of the interpolated values of stellar mass, for each star we also generated 1000 test values of $V_{0}$ and $(V-I)_{0}$ adopting a normal distribution centered on the nominal values and with a width equal to the photometric errors. We then calculated the mass corresponding to each pair of simulated values of $V_{0}$ and $(V-I)_{0}$ and adopted as uncertainty the size of the resulting range of mass. We also included in our analysis the individual effective temperature and gravity index $\gamma$ \citep{Damiani2014AA.566A.50D} obtained from the fourth internal data release GES campaign, published by \citet{Venuti2018AA.609A.10V}. Gaia-ESO Survey data are available for 1892 stars in the NGC 2264 field, over 80\% of which were observed using the GIRAFFE instrument mounted on the FLAMES spectrograph at the Very Large Telescope (VLT). Among the 44 stars in the three correlation groups, GES provides measurements of T$_{\rm eff}$ for 38 stars and of $\gamma$ for 36 stars. \par
        
        \begin{figure*}[]
        \centering      
        \includegraphics[width=8cm]{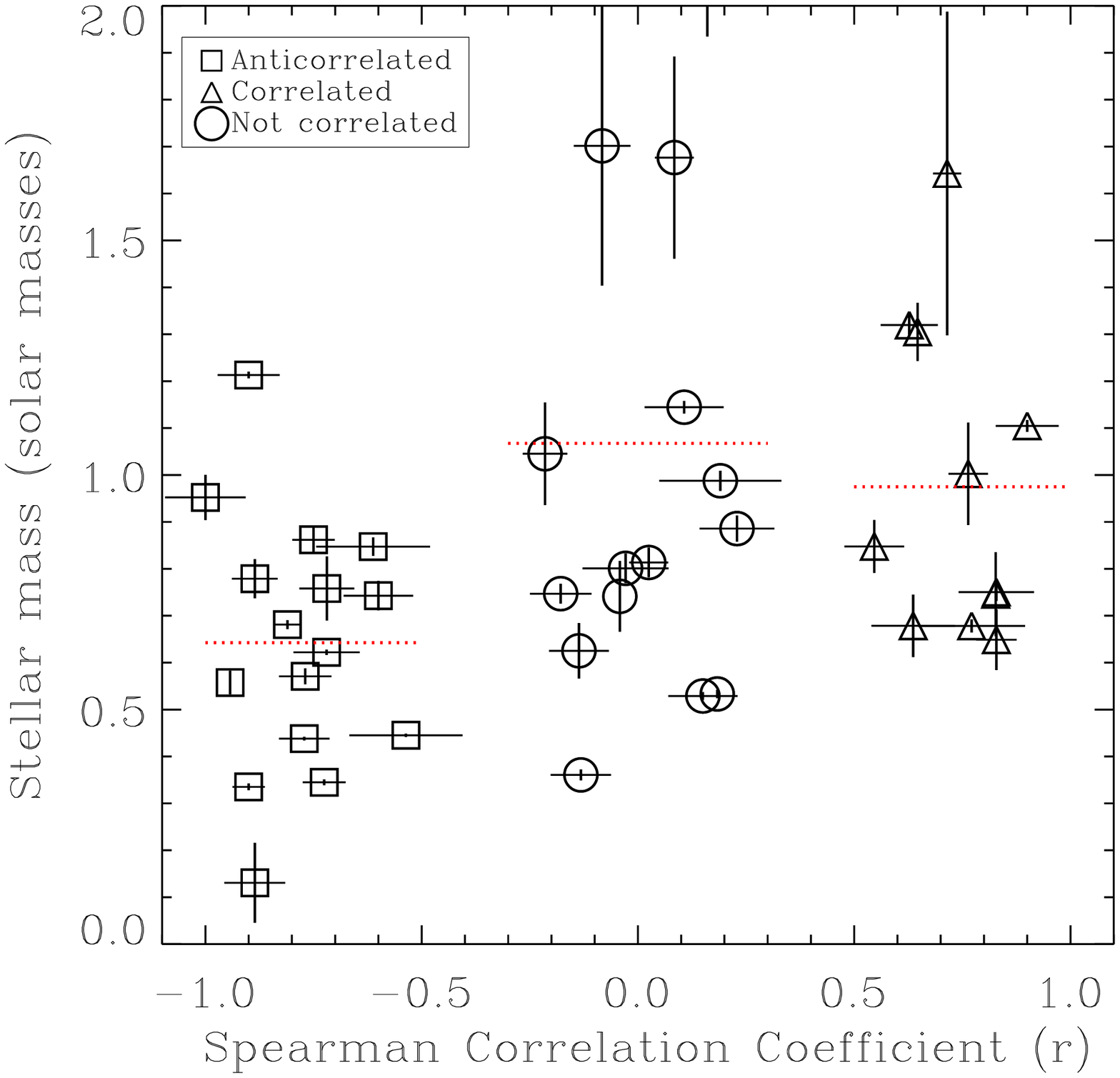}
        \includegraphics[width=8cm]{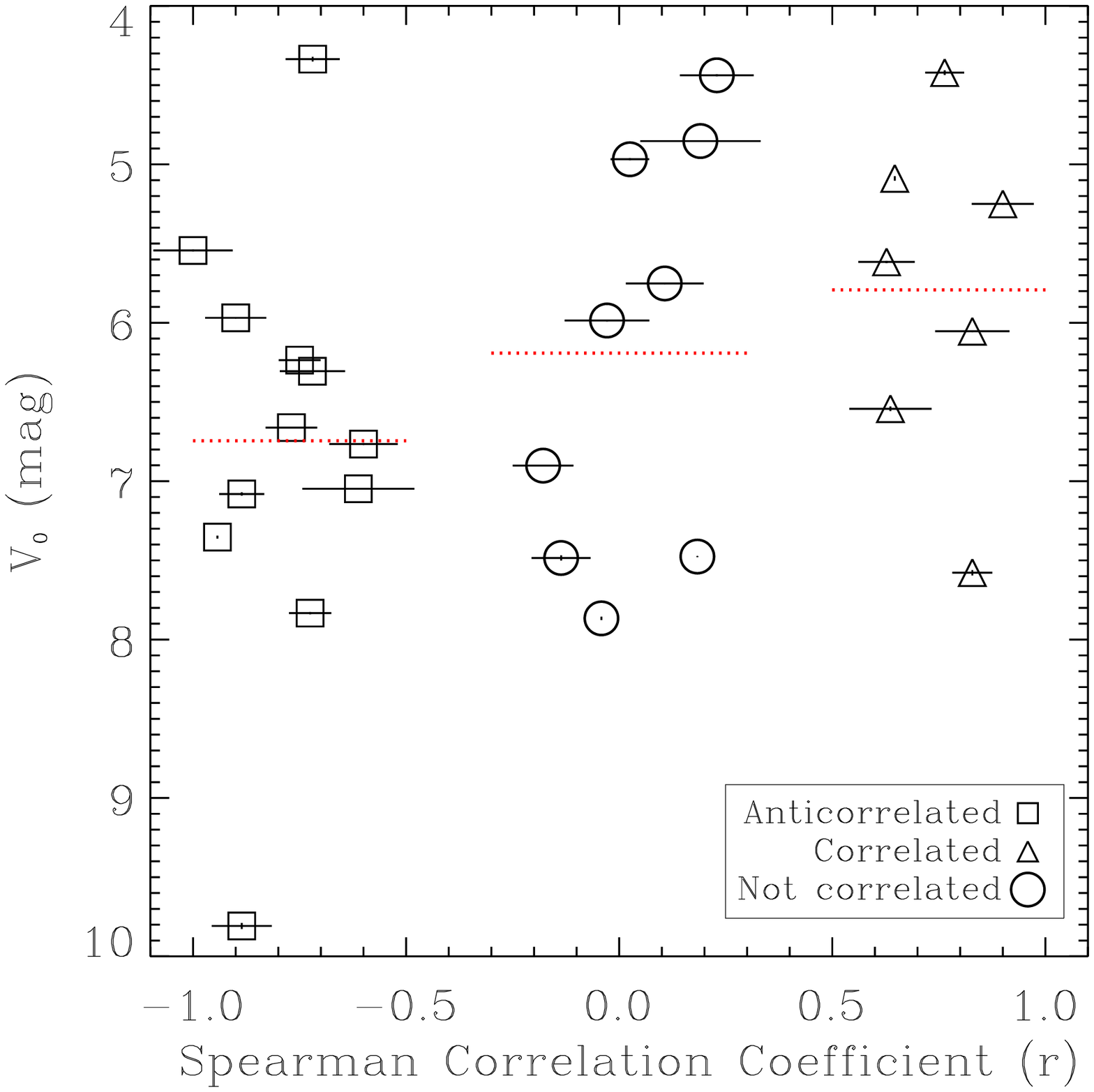}
        \includegraphics[width=8cm]{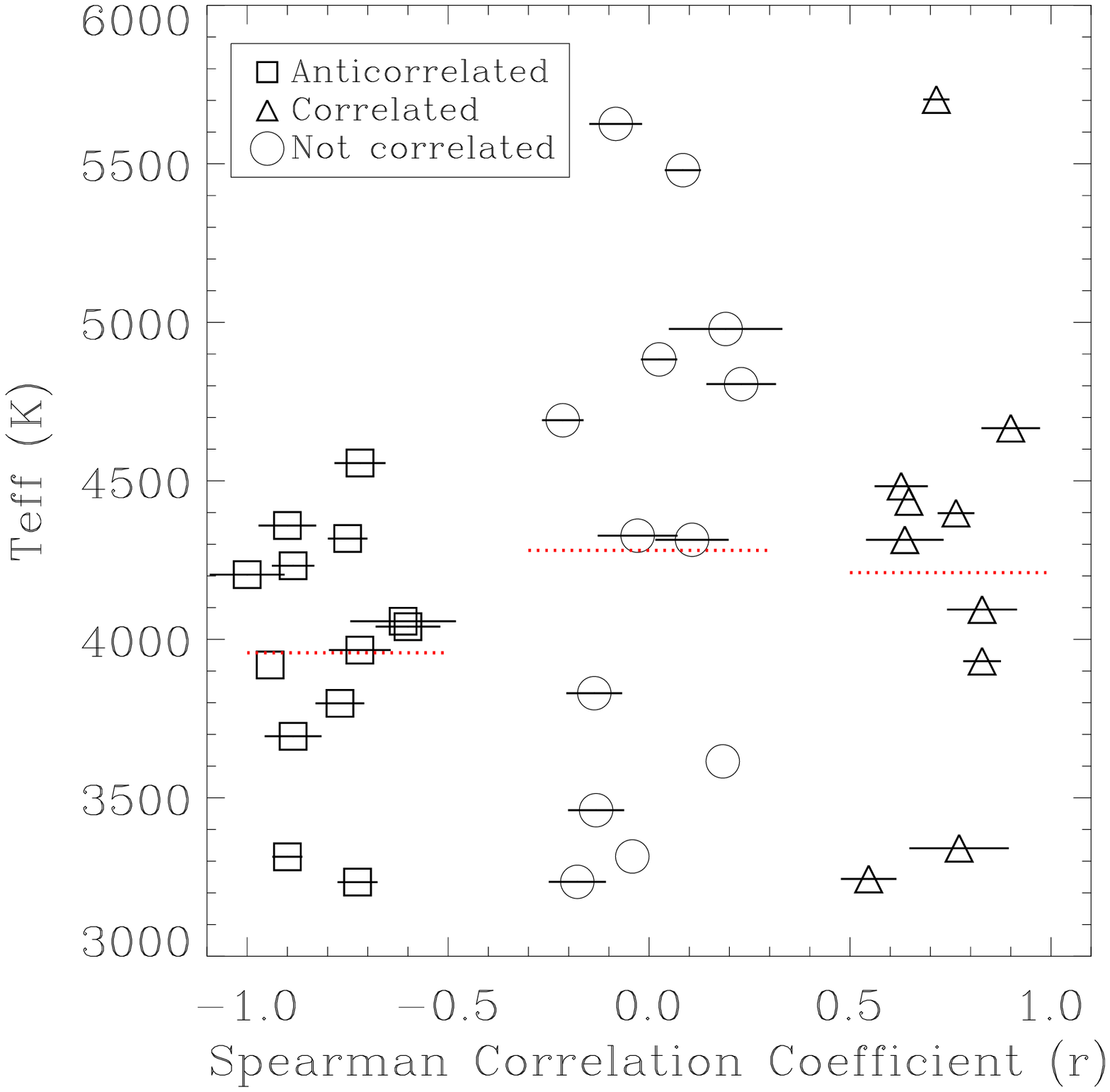}
        \includegraphics[width=8cm]{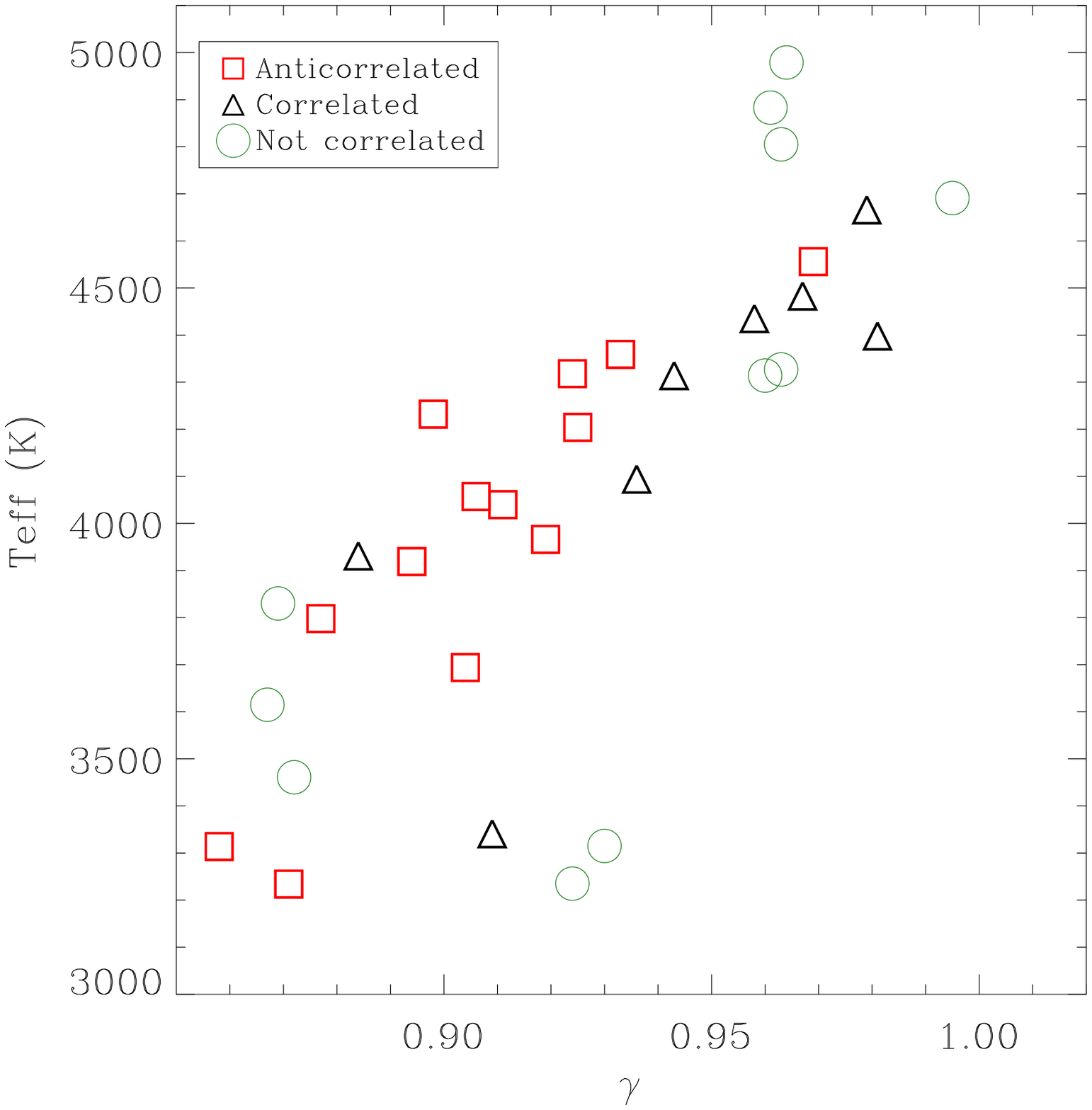}
        \caption{Stellar masses (top left), $V_{0}$ (top right), and T$_{\rm eff}$ (bottom left) vs. the Spearman's correlation coefficient. In the bottom right panel we also show the T$_{\rm eff}$ vs. $\gamma$ scatter plot. Stars with different correlation between the optical and X-ray variability are marked with different symbols as shown in the labels. The horizontal dotted lines mark the median values. All the plots suggest that the stars in the ``anti correlated'' group are on average less massive than those in the other two groups.}
        \label{mass_fig}
        \end{figure*}
        
        The individual masses, $V_{0}$ magnitudes, and effective temperatures of the stars in the three correlation groups, as well as the T$_{\rm eff}$ versus $\gamma$ scatter plot of the stars in our sample, are shown in Fig. \ref{mass_fig}. These figures suggest that the stars in the anti-correlated groups are on average less massive than those in the other two groups, with a significant population below 0.6-0.7$\,$M$_{\odot}$ which is missing among the correlated and not correlated stars. A two-sided Kolmogorov-Smirnov test between the distribution of masses of the anti-correlated versus correlated and not correlated stars together marginally confirms this difference (the resulting KS statistic is 0.4, with a significance level of 5.4\%).

\subsection{Stellar rotation}
\label{rotation_sect}

\begin{figure}[]
        \centering      
        \includegraphics[width=8cm]{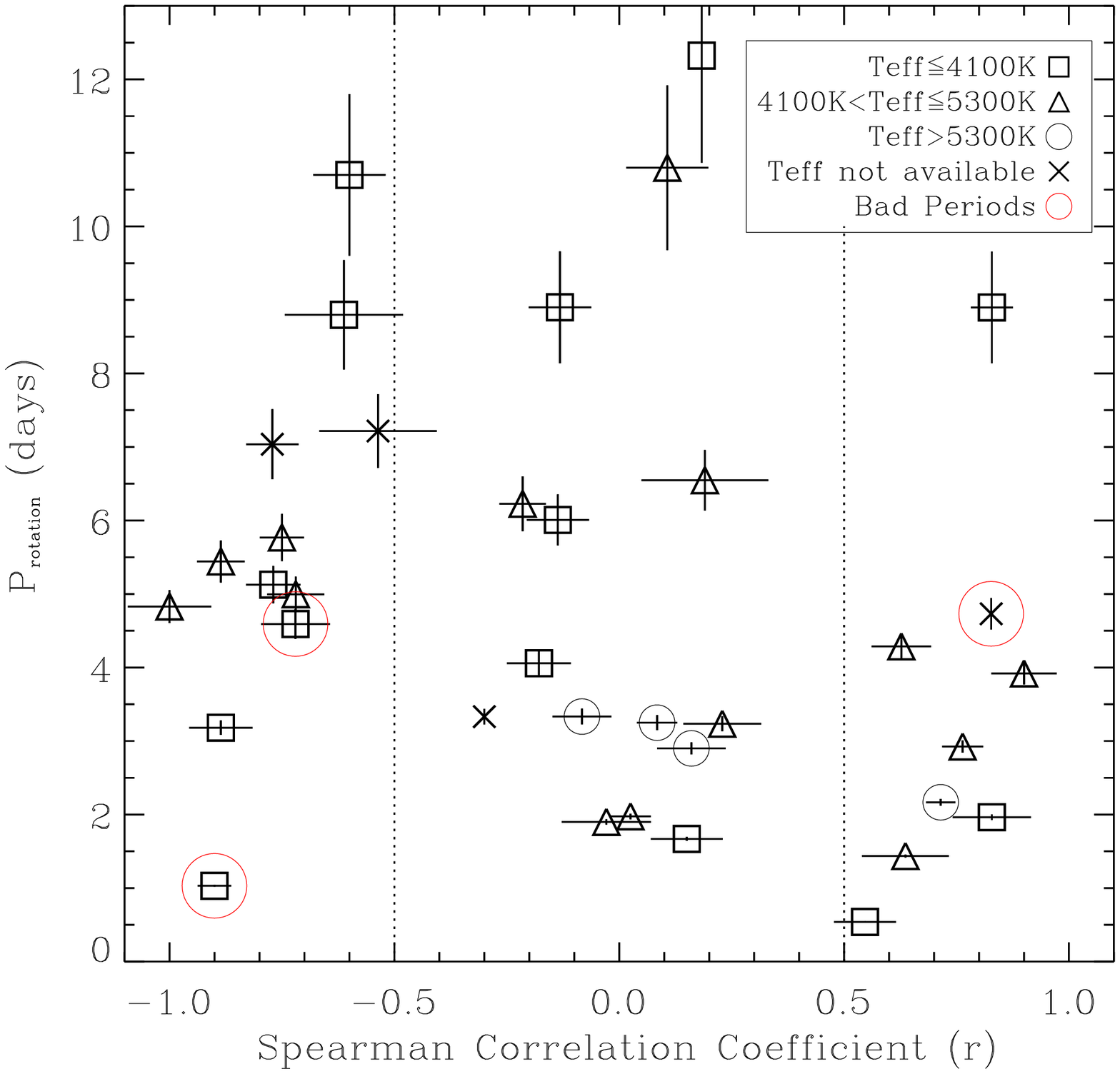}
        \caption{Rotation period vs. Spearman's correlation coefficient. Stars are sampled in three ranges of T$\rm_{eff}$ and plotted with different symbols, as shown in the label. Stars whose period is not well determined are marked with a red circle. anti-correlated stars are on average slower rotators than the correlated stars.}
        \label{corr_per}
        \end{figure}

The rotation periods of the stars in the three groups are shown in Fig. \ref{corr_per}. Rotation periods were calculated by \citet{Venuti2017AA.599A.23V} and \citet{Affer2013MNRAS.430.1433A} from the CoRoT light curves. We also marked stars whose periods are not well determined\footnote{The anti-correlated source with the smallest rotation period is Mon-220, the only multi-periodic star in the three groups, with a primary rotation period equal to 0.75$\,$days and a secondary period of 7.46$\,$days}. Stars with correlated and anti-correlated optical and X-ray flux variability have different rotational properties: anti-correlated stars are in fact slower rotators, with all but one star with a reliable estimate of the rotation period having a period in the 4--12 days range, while all correlated sources but one have P$\rm_{rot}\leq 5\,$days. This difference is confirmed by a K-S test on the two distributions, yielding a KS statistic of 0.91 and a significance level of $10^{-5}$\%. Stars in the not correlated group have rotation periods covering the whole 1--12 days range. As in Fig. \ref{fxvsr}, we plotted stars in three different ranges of effective temperature with different symbols, roughly corresponding to M, K, and G stars. No evident pattern with the stellar spectral type is observed in Fig. \ref{corr_per}. \par

\subsection{Peak-to-peak and color optical variability}
\label{amplitude_sect}

    It is intuitive that the peak-to-peak optical amplitude variability is connected with the faction of stellar photosphere covered by cold spots and to the evenness of the spot distribution. In order to calculate the peak-to-peak variability in each star, we first split each full 40-day CoRoT light curve into an integer number of rotation periods. We then fitted each portion of the light curve with a fifth-degree polynomial and calculated the amplitude as the normalized difference between maximum and minimum of the best-fit curve. We then associated with each star the median value calculated over all the portions of the light curve.  We use the best-fit polynomial to limit the effect of statistical fluctuations of the light curve or by unresolved small flares on our estimated amplitude. 

        \begin{figure}[]
        \centering      
        \includegraphics[width=8cm]{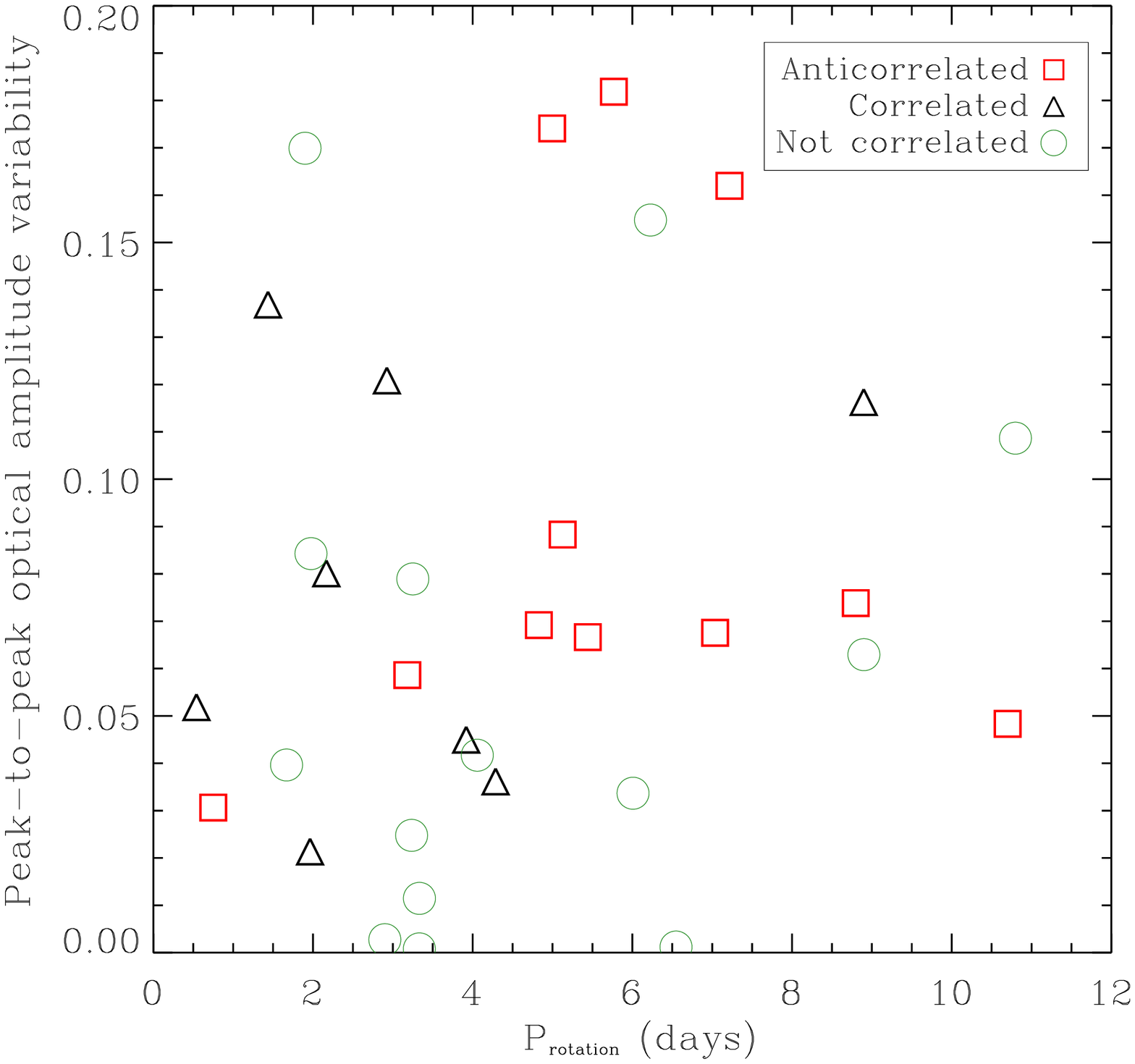}      
        \caption{Peak-to-peak optical amplitude variability as a function of the rotation period. Stars are marked with symbols accordingly with their correlation class.}
        \label{amp_fig}
        \end{figure}

        Comparing the distribution of peak-to-peak amplitude variability in the three correlation classes, the most evident difference is that the not correlated sources have a smaller median value ($\sim$0.04 to be compared with $\sim$0.07 for the anti-correlated and $\sim$0.08 for the correlated sources). Differences can also be observed in the peak-to-peak amplitude variability versus rotation period scatter plot shown in Fig. \ref{amp_fig}, where we note that the anti-correlated sources populate mainly the upper portion of the plot (i.e., 9 out of 11 stars have peak-to-peak amplitude variability larger than 0.05), while 10 of the 14 not correlated sources populate the lower left portion of the diagram (i.e., P$\rm_{rotation}\leq$6$\,$days and peak-to-peak amplitude variability smaller than 0.1).\par
        
        We also took advantage of the fact that most of the stars in the three groups have been observed in $u$ and $r$ bands with the CFHT using the wide-field camera MegaCam, and that \citet{VenutiBIS2015AA} provided their variability in these two bands and in $u-r$. The analysis of this color is useful since different color variability in the stars in the three correlation groups may reflect different temperature contrast between the quiet photosphere and the surface magnetic activity mainly responsible for the observed variability (i.e., spots and/or faculae). In the Sun, for instance, a larger 
        temperature contrast is observed between the quiet photosphere and spots rather than faculae \citep{FrohlichLeanAARv.12.273F,RodriguezLedesma2009AA.502.883R}. 

        \begin{figure}[]
        \centering      
        \includegraphics[width=7cm]{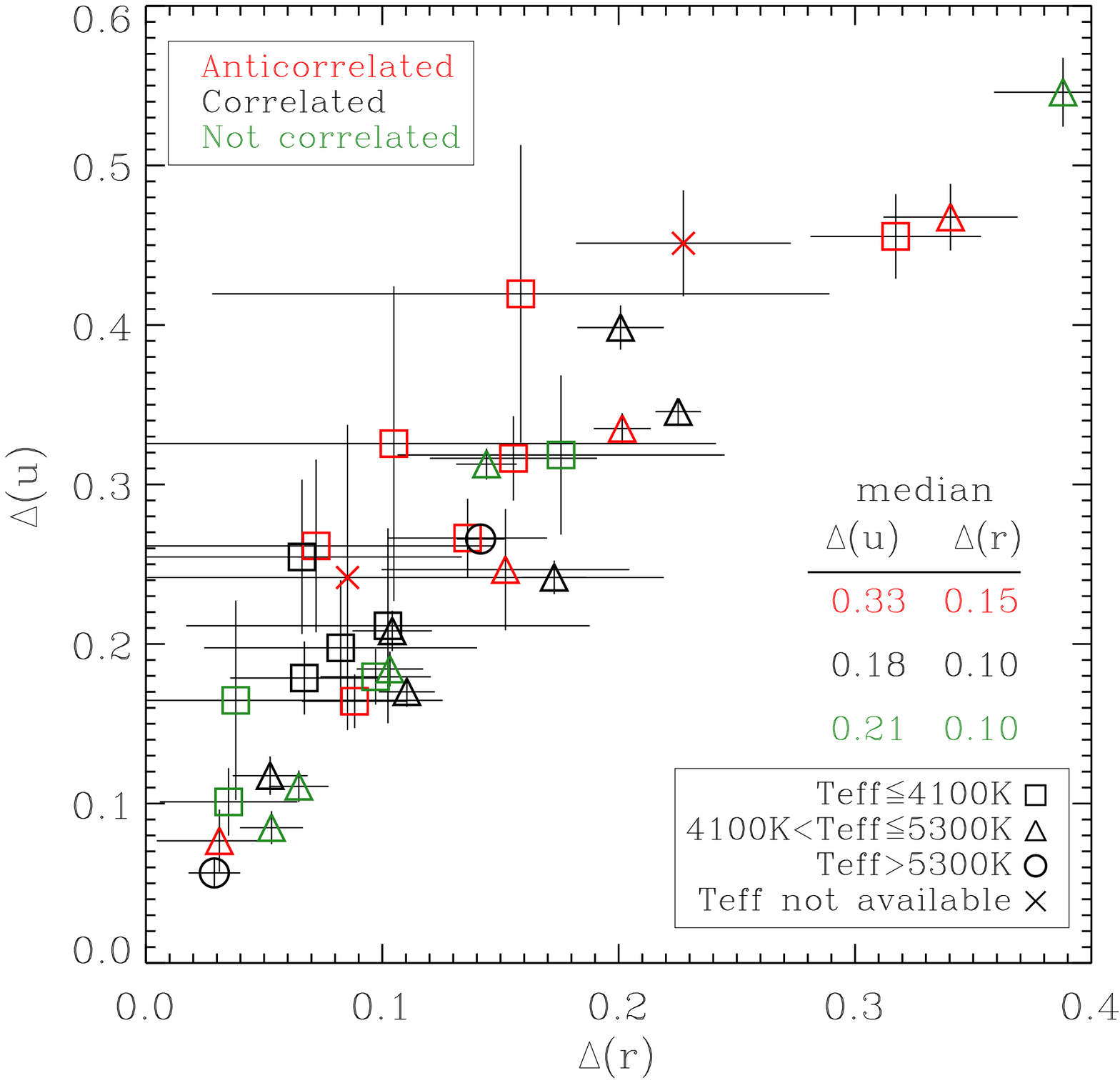}
        \includegraphics[width=7cm]{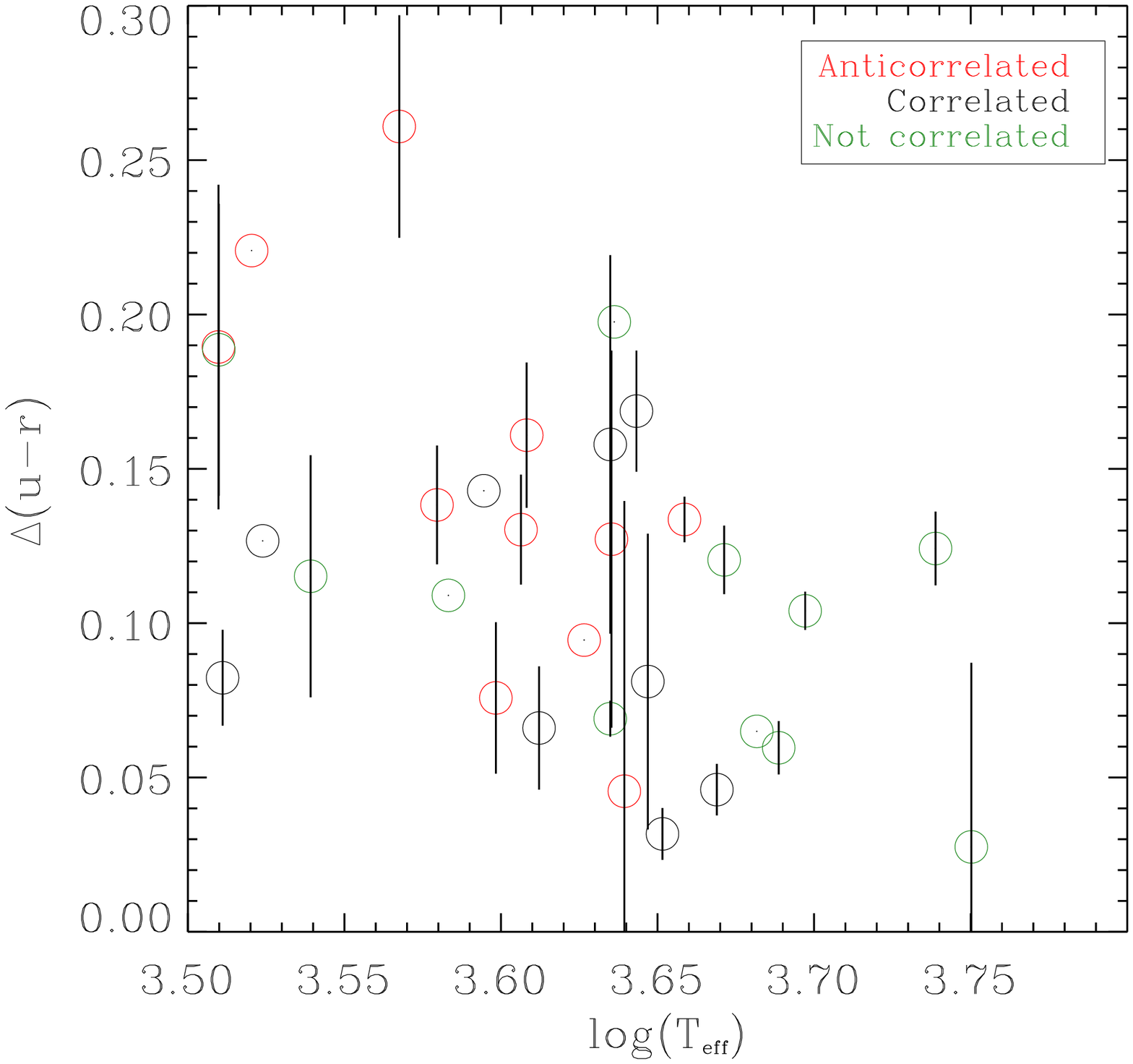}
    \includegraphics[width=7cm]{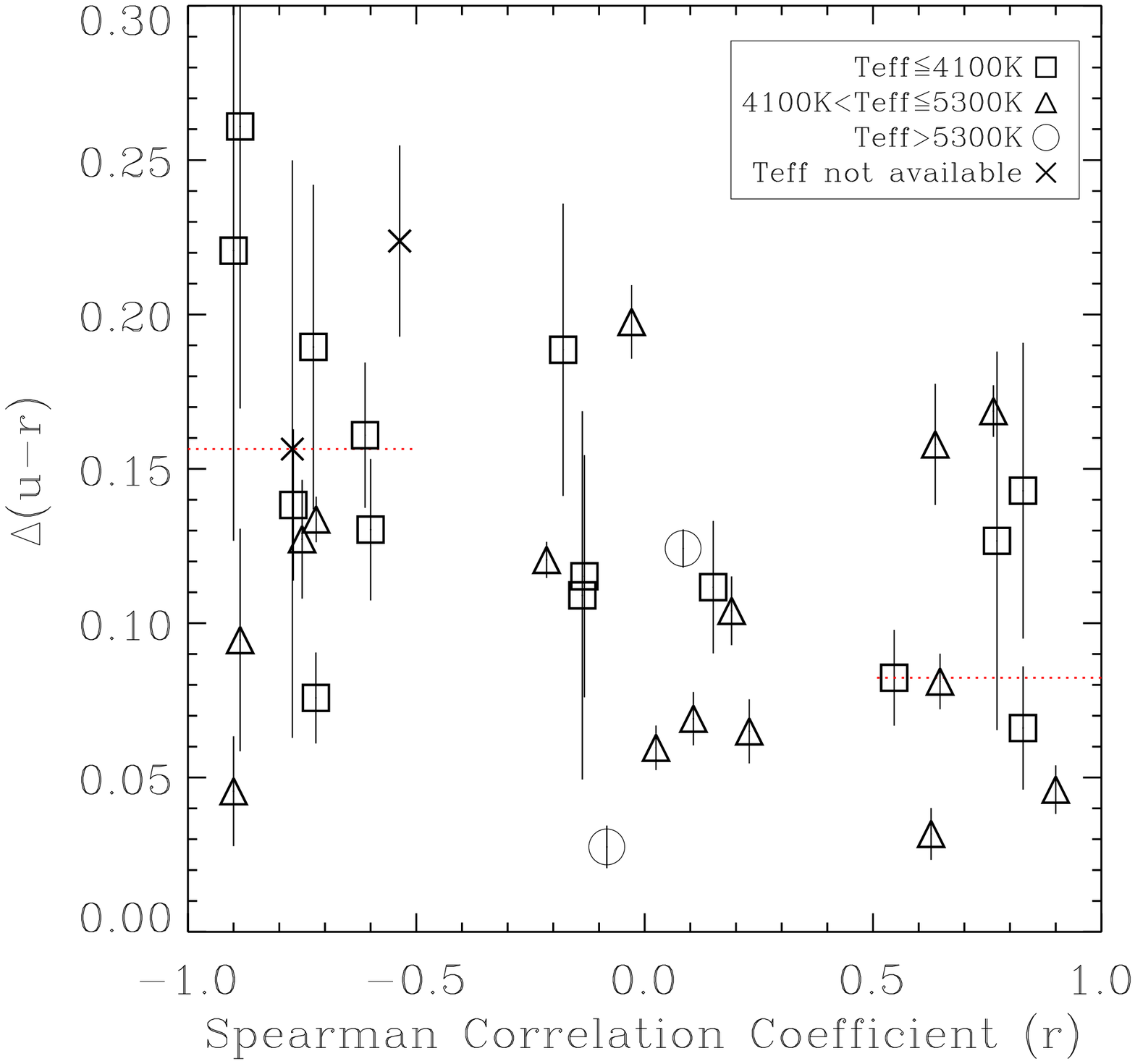}  
        \caption{Top panel: Variability in $u$ and $r$ bands for the stars in the three correlation groups. Central panel: Variability in $u-r$ vs. stellar effective temperature; Bottom panel: Variability in $u-r$ vs. the Spearman correlation coefficient. In the top and bottom panels different symbols are used to mark stars in three different ranges of effective temperature, roughly corresponding to M, K, and G stars. In the top and central panels, different colors are used to plot stars in the three correlation groups. Despite the large spread observed, the plots suggest a slightly larger median variability in color for the anti-correlated sources.}
        \label{col_fig}
        \end{figure}
        
        The $\Delta(u)$ versus $\Delta(r)$, $\Delta(u-r)$ versus T$\rm_{eff}$, and $\Delta(u-r)$ versus the Spearman correlation coefficient scatter plots for the stars in the three correlation groups are shown in Fig. \ref{col_fig}. The large spread observed in these plots prevents us from making strong statements on the differences between stars in the three correlation groups, however a slightly larger color and $u$ band variability is observed in the anti-correlated sources, and the color variability declines with the stellar effective temperature. This can likely be due to a larger temperature difference between the quiet photosphere and the dominant surface magnetic activity for stars in the anti-correlated group and for cool stars in our sample. The Spearman correlation test run over the $\Delta(u-r)$ vs. T$\rm_{eff}$ scatter plot results in a correlation coefficient $r$=-0.5 with P($r$)=0.003. \par

\section{The physics behind the correlation between optical and X-ray variability}
\label{discuss_sect}

    The optical variability induced by the rotational modulation of magnetic spots is due to the temperature difference between spots and quiet photosphere. In T~Tauri stars, the difference is between $\sim$500$\,$K and $\sim$1000$\,$K. For instance, the temperature difference between quiet photosphere and spots in the M dwarf V374~Peg (mass 0.28$\,$M$_{\odot}$, radius 0.28$\,$R$_{\odot}$, rotation period 0.446$\,$days) is $\sim$400$\,$K \citep{Morin2008MNRAS.384.77M}, while a larger difference ($\sim$1000$\,$K) has been observed in the K0-K2 star AB~Dor, which has a similar mass and radius to our Sun and a rotation period of 0.514$\,$days \citep{Donati1999MNRAS.302.437D}. Additionally, spots can last several stellar rotation periods, resulting in a stable modulation of the optical emission. Anti-correlated optical and X-rays variability can therefore be understood in terms of rotational modulation of dark spots in the photosphere which emerge on the line of sight simultaneously with coronal active regions bright in X-rays. As we know from the Sun,  cold spots are characterized by intense magnetic fields (2-3$\,$KG) and they typically spatially correspond to X-ray-bright coronal active regions.  \par
    
    Similarly, correlated optical and X-ray variability can result from the optical brightening due to a photospheric network of faculae whose brightening is dominant over the darkening due to photospheric spots, with again these features being spatially coincident with coronal active regions. We know from our Sun, which is orders of magnitude less active than young PMS stars, that faculae can result in a significant brightening. In fact, the solar total irradiance increases by about 0.1\% during the peak of solar activity, when the spot coverage is the largest, rather than decreasing, because of the presence of a bright network of faculae in the photosphere. Therefore, the most natural interpretation of these two temporal behaviors is that anti-correlated sources are spot-dominated stars, while the correlated sources are faculae-dominated. The larger median amplitude variability observed in $u-r$ for the anti-correlated (Fig. \ref{col_fig}) sources could reflect the larger temperature contrast that exists between the quiet photosphere and cold spots in the anti-correlated sources than with faculae in the correlated sources. In this picture, the observed anti-correlation between $\Delta(u-r)$ and stellar effective temperature shown in Fig. \ref{col_fig} also suggests that stellar spectral type is an important factor that dictates the dominant surface stellar magnetic activity: with cool stars being typically spot-dominated, while the effects of faculae is more important for hot stars. Another possible important factor is the distribution of spots over the surface: a more even distribution of spots results in a smaller $u-r$ variability than that induced by few large and well-localized spots. \par

    \citet{Shapiro2016AA.589A.46S} quantified the contribution of faculae and spots to the solar variability as it would be observed with Kepler at different inclination angles. They have found that spots dominate the variability of solar and solar-like stars  for low and intermediate inclinations, while faculae dominate at very high inclinations. This finding suggests that if a photospheric active region rich in spots and faculae is observed during the whole stellar rotation, the effect of faculae may dominate when the active region is observed close to the edge of the stellar disk, while the effect of spots dominates during the rest of the rotation phase. Therefore, despite the fact that we analyze in our paper a wider range of spectral types than that analyzed by \citet{Shapiro2016AA.589A.46S}, the classification of a star as an anti-correlated or correlated source could be affected by the way the light curve is sampled next to its minima and peaks. For instance, if the number of intervals next to the minima is dominant over the number next to the peaks, this could artificially enhance the effect of the faculae close to the stellar edge compared to the effect of the spots across the rest of the stellar disk. To test whether this may occur in our stars, we chose to analyze in detail the “anti-correlated” source Mon-198. This source is bright enough in X-rays to allow us to sample the light curve with N$\rm _{phot}$=20 over just the optical minima and maxima, and to perform in both cases a correlation test with high confidence. We verified that the classification of Mon-198 as an anti-correlated source holds even by restricting the correlation tests to time intervals close to the optical minima or to the maxima, and thus the correlation does not depend on the phase at which the correlation is observed. This therefore suggests that the sampling does not affect the classification of sources in the correlated and anti-correlated groups, although a more detailed theoretical analysis of the importance of spots and faculae observed at different inclinations or close to the edge of the stellar disk in active T~Tauri stars would be necessary to firmly address this point.
    
    Figure \ref{corr_per} shows that stars with anti-correlated optical and X-ray variability, which we suggest are spot-dominated sources, are predominantly slow rotators compared to stars with correlated variability, which can be faculae-dominated sources. This suggests that the relative importance of spots and faculae is also related to stellar rotation. This is known to occur in main sequence stars. In fact, young main sequence stars are on average faster rotators than old main sequence stars, resulting in more intense magnetic activity in the former stars than in the latter \citep[e.g.,][]{PallaviciniGRV1981}. Evidence exists showing that younger and more rapidly rotating main sequence stars are typically spot-dominated while older and slower rotating active stars are typically faculae-dominated. This has been suggested, for instance, by \citet{RadickLSB1998ApJS} and \citet{LockwoodSHH2007ApJS}, who compared long-term (over a baseline longer than 1 year) and short-term (over a baseline shorter than 1 year) variability of the Sun and 35 Sun-like stars using optical photometric data to mark the variability of photospheric emission, and spectroscopic observations of the Ca~II~H+K lines to characterize the variability of chromospheric emission. Other studies, such as \citet{RadickLB1990} and \citet{BoisseBS2012}, support the evidence that active main sequence stars are typically spot-dominated. \par

    Our study suggests that over a time-period of hours to a few days, PMS stars behave differently from MS stars observed over longer time baselines, with the slow rotators (P$\rm_{rot}\geq4.5\,$days), namely 7 M and 7 K stars, being mainly spot-dominated and with anti-correlated optical and X-ray variability, while fast rotators (P$\rm_{rot}$$\leq$$4.5\,$days), namely 5 M, 7 K, and 4 G stars, are mainly faculae-dominated with correlated optical and X-ray variability. This difference is not surprising given that other existing studies indicate that the relations between activity, age, and stellar rotation valid for main sequence stars are not valid for young and very active stars, and it may highlight a further difference existing between stars in these two evolutionary stages. For instance, while young and active MS stars are typically spot-dominated, correlated chromospheric and photospheric variability was observed in the young K2V~LQ~Hya star, with a rotation period of 1.6 days \citep{StrassmeierRWH1993} and in the active and rapidly rotating binary star TZ~CrB \citep[F6V+G0V, P$\rm_{orb}$=1.1 days;][]{FrascaCM1997}. On the other hand, no variability correlation is observed in two rapidly rotating Pleiades stars studied by \citet{Stout-BatalhaVogt1999}. \par

        To interpret the observed connection between stellar rotation and the simultaneous optical and X-ray variability (Fig. \ref{corr_per}), we first note that T~Tauri stars can be characterized by complex magnetic fields whose morphology departs significantly from a simple dipolar field \citep{JardineCDG2006,GregoryDonati2011AN.332.1027G}. In a sample of ten T~Tauri stars, including well-studied objects such as AA~Tau, BP~Tau, and TW~Hya, \citet{Johnstone2014MNRAS.437.3202J} found that the topology of the magnetic field is connected to stellar rotation, with the rapid rotators hosting preferentially weak and complex magnetic fields, while slow rotators have fields with a dominant dipolar topology. If the rotation is related to the geometry of the magnetic field, then it may be related also to the dominant surface magnetic activity. \citet{RodriguezLedesma2009AA.502.883R} further suggested that less complex (mainly dipolar) magnetic fields typically result in a less numerous population of spots, which are typically large and isolated, while a more complex field produces a larger coverage by a population of typically small stellar spots. This could be the explanation for the different rotational properties among the stars in the correlated and anti-correlated groups: with the latter stars being spot-dominated and on average slow rotators, characterized by an ordered magnetic field producing preferentially few and large stellar spots; while stars in the correlated group are on average fast rotators, with a complex magnetic fields producing preferentially a rich population of small spots and resulting in faculae-dominated sources with correlated optical and X-rays variability. Despite the differences between main sequence and T~Tauri stars described above, this scenario may also apply to the solar cycle, with the solar magnetic field being mainly dipolar during the minimum of the activity cycle. On the other hand,  during the maximum the magnetic field becomes more complex, resulting in a richer spot population and in a total irradiance variability which is dominated by the brightening due to the faculae. However, the situation is likely more complex than that described in this scenario. First, ordered dipolar large-scale magnetic field can still hold significant energy in small-scale field, which is mainly responsible for the surface magnetic activity. This has been suggested to be important for low-mass fully convective stars by \citet{ReinersBasri2009AA.496.787R}. Second, the stars analyzed by \citet{Johnstone2014MNRAS.437.3202J} are disk-bearing stars, while our work focuses on diskless stars. Recent results from the MaTYSSE project\footnote{https://matysse.irap.omp.eu/doku.php} (P.I.: Donati) show that the magnetic fields of weak-line T Tauri stars (WTTS) cover a wide range of topology, going from simple dipolar fields to more complex geometries, with no significant correlation between the complexity of the magnetic field and the stellar parameters such as rotation \citep{Hill2019MNRAS.484.5810H}. Considering both diskless and disk-bearing stars, the only connection between the morphology of the magnetic field and stellar properties supported by MaTYSSE results so far is the evidence that stars with T$\rm_{eff}$$>$4300$\,$K have a more complex field than cooler stars. This may be part of the reason why the anti-correlated group have a slightly smaller content of stars with T$\rm_{eff}$$>$4100$\,$K (4/12) than the correlated group (5/9) and why we observe a larger variability in $u-r$ color for cool stars compared with hot stars (Fig. \ref{col_fig}) in the scenario in which an ordered magnetic field is more typical of spot-dominated sources (which are expected to have larger variability in $u-r$) while a complex field is typical of faculae-dominated sources.  \par

        The plots shown in Fig. \ref{mass_fig} also suggest that in our sample, spot-dominated sources have typically lower mass than faculae-dominated sources. This does not imply that in NGC~2264 slow rotators have lower masses than rapid rotators. On the contrary, \citet{Venuti2017AA.599A.23V} presented a detailed analysis of stellar rotation in NGC~2264 using CoRoT data and found marginal evidence that low-mass objects typically rotate faster than high-mass stars (even if their analysis does not allow them to reject the hypothesis that the observed period distribution does not depend on stellar mass). More convincing evidence on a typical faster rotation of low-mass objects in the PMS phase was recently obtained in Upper~Sco \citep{Rebull2018AJ.155.196R} and Taurus (L. Rebull, private communication) from Kepler/K2 data. \citet{Venuti2017AA.599A.23V} also found no clear connection between the rotational properties of the stars in NGC~2264 and their internal structure, likely because this stellar population is too young to result in a significant development of a radiative core in the observed mass range. The finding that anti-correlated sources are on average less massive than stars in the other two groups is however quite interesting, since it indicates that stellar rotation is not the only factor dictating the main surface stellar magnetic activity in such a coeval young stellar population, with stellar mass also playing an important role. This could happen because of a rotation--mass relation, or because of the mass-dependent evolution of the internal structure of the star, which is crucial for the dynamo process and the stellar magnetic properties  \citep[e.g.,][]{DonatiLandstreet2009ARAA.47.333D}. The possibility that the dominance of the importance of faculae over that of spots in stars decreases with stellar mass has also been predicted by the MURAM code \citep[e.g.,][]{Beeck2015AA.581A.43B}. It might also be that, because of the limiting magnitudes of the CoRoT data, we only observed the younger (and therefore brighter) low-mass stars, which may also be the slowest rotators in our sample since they had less time to spin up after their disks were dispersed.
        
        We collected some evidence supporting the idea that a larger number of not correlated sources are more active in X-rays than those in the other two groups (Fig. \ref{fxvsr}). The brightness distribution of stellar coronae is typically clumpy depending on how the active regions are distributed \citep[e.g.,][]{Gregory2006MNRAS.373.827G}. This results in a rotationally modulated X-ray emission observed, for instance, in the T~Tauri stars of the Orion Molecular Cloud \citep[][]{FlaccomioMSF2005ApJ} as part of the \textit{Chandra} Orion Ultradeep Project \citep{GetmanFBM2008ApJ}. The lack of correlation in most of the stars in the not correlated sample could be due therefore to a more uniform brightness distribution of coronae due to a higher coronal activity, compared to stars in the other two correlation groups, resulting in a weak rotational modulation of X-ray emission. In principle, no correlation between optical and X-ray emission is expected also for stars observed along a line of sight which is parallel to the rotation axis. This could be the case for the not correlated stars with peak-to-peak amplitude variability being almost zero in Fig. \ref{amp_fig}. It is also likely that sampling of the light curve forced the classification of some sources as  not correlated. This could be the case of stars like Mon-223 (Fig. \ref{variab_others_1}), for which we have been able to sample only the minima of the optical light curve.\par

\section{Conclusions}
In this paper we analyze the simultaneous optical (from CoRoT) and X-ray (from $Chandra$/ACIS-I) variability in 74 pre-main sequence stars of NGC~2264 without circumstellar disks or with a transitional disk, as part of the project CSI~2264 (Coordinated Synoptic Investigation of NGC~2264). We selected 16 stars with anti-correlated X-ray and optical simultaneous variability, 11 stars with correlated variability, and 17 stars with uncorrelated variability. \par

    Our analysis suggests that stars with anti-correlated optical and X-ray variability are spot-dominated stars, where stellar rotation modulates the optical fading due to large photospheric spots emerging simultaneously with coronal active regions bright in X-rays; while sources with correlated variability are faculae-dominated stars, with the optical brightening due to the network of faculae being dominant over the optical fading due to stellar spots. This is marginally supported by the larger median amplitude variability in $u-r$ observed in the former stars, compatible with a larger temperature difference between the main magnetic phenomenon (spots or faculae) with respect to the quiet photosphere. In some of the stars with uncorrelated optical and X-ray variability, the lack of correlation is likely due to a larger coverage by active regions of stellar coronae, resulting in a X-ray variability which is more stochastic than what would be expected if it were modulated by stellar rotation. Alternatively, some of the stars in this group show a nearly zero peak-to-peak optical amplitude variability. For these stars, it is more likely that the line of sight is almost parallel to the rotation axis, resulting in a very low rotational modulation. 
    
We explored the differences in stellar properties among the stars in the three correlation groups. For instance, stars with anti-correlated optical and X-ray variability rotate slower than those with correlated variability. In all of the former stars but one,  P$\rm_{rotation}\geq\,$4.5$\,$days, while in all the latter stars but one P$\rm_{rotation}\leq\,$6$\,$days. This difference is confirmed by a K-S test. We also found that the amplitude of $u-r$ color variation is larger in cool stars than in hot stars and is larger for stars in the anti-correlated group than in the correlated. Furthermore, we observe a slightly larger content of KG stars compared with M stars in the correlated stars compared to the `anti-correlated'' group, and that stars with anti-correlated variability are on average less massive than those in the other two groups. It is likely that the dominant magnetic surface activity (spot or faculae) is thus related in a complex way to a combination of stellar properties such as mass and rotation.  Conversely, as proposed by previous studies, the dominant magnetic activity could be related to the topology of the magnetic field, with complex magnetic fields producing a larger population of small spots than simple dipolar fields which preferentially produce few large spots. However, as supported by the results obtained by the MaTYSSE project so far, the situation should be more complicated than this, since no clear relation between the morphology of the magnetic field and stellar parameters has been confirmed to date by this project. 

Our paper also shows how studies of the simultaneous variability in optical and X-rays can help to shed light on the stellar magnetic activity and can provide important information on the evolution of stars if applied to rich stellar samples of different age.

\begin{acknowledgements}
M. G. G. acknowledges partial support from the agreement ASI-INAF n. 2017-14-H.0. The scientific results reported in this article are based on observations made by the Chandra X-ray Observatory and the CoRoT satellite. The authors of this paper also made an extensive use of the Xspec and TOPCAT softwares and the NASA's Astrophysics Data System. Finally, we wish to thank the anonymous reviewer for valuable insight into the paper and important suggestions.
\end{acknowledgements}
% \clearpage

% \pagebreak
%\addcontentsline{toc}{section}{\bf Bibliografia}
\bibliographystyle{aa}
\bibliography{biblio}

\end{document}